\documentclass[twocolumn,tighten,twocolappendix]{aastex631}

\received{January 4, 2022}
\revised{March 14, 2022}
\accepted{March 17, 2022}
\submitjournal{ApJ}

\usepackage{url, soul, bm}
\usepackage{amsmath}      

\shorttitle{Trifid North}
\shortauthors{Kuhn et al.}

\graphicspath{{./}{figures/}}

\begin{document} 

\title{The Effect of Molecular Cloud Properties on the Kinematics of Stars Formed in the Trifid Region}

\correspondingauthor{Michael A. Kuhn}
\email{mkuhn@astro.caltech.edu}

\author[0000-0002-0631-7514]{Michael A. Kuhn}
\affil{Department of Astronomy, California Institute of Technology, Pasadena, CA 91125, USA}

\author{Lynne A. Hillenbrand}
\affil{Department of Astronomy, California Institute of Technology, Pasadena, CA 91125, USA}

\author[0000-0002-5077-6734]{Eric D. Feigelson}
\affil{Department of Astronomy \& Astrophysics, Pennsylvania State University, 525 Davey Laboratory, University Park, PA 16802, USA}

\author[0000-0002-4958-8648]{Ian Fowler}
\affil{Department of Astronomy, California Institute of Technology, Pasadena, CA 91125, USA}

\author[0000-0002-6137-8280]{Konstantin V. Getman}
\affil{Department of Astronomy \& Astrophysics, Pennsylvania State University, 525 Davey Laboratory, University Park, PA 16802, USA}

\author[0000-0002-7872-2025]{Patrick S. Broos}
\affil{Department of Astronomy \& Astrophysics, Pennsylvania State University, 525 Davey Laboratory, University Park, PA 16802, USA}

\author[0000-0001-9062-3583]{Matthew S. Povich}
\affil{Department of Physics and Astronomy, California State Polytechnic University, 3801 West Temple Ave., Pomona, CA 91768, USA}

\author[0000-0002-1650-1518]{Mariusz Gromadzki}
\affil{Astronomical Observatory, University of Warsaw, Al. Ujazdowskie 4, 00-478 Warszawa, Poland}

\begin{abstract}
The dynamical states of molecular clouds may affect the properties of the stars they form. In the vicinity of the Trifid Nebula ($d=1180\pm25$~pc), the main star cluster (Trifid Main) lies within an expanding section of the molecular cloud;  however, $\sim$0.3$^\circ$ to the north (Trifid North), the cloud's velocity structure is more tranquil. We acquired a Chandra X-ray observation to identify pre-main-sequence stars in Trifid North, complementing a previous observation of Trifid Main. In Trifid North, we identified 51 candidate pre-main-sequence stars, of which 13 are high-confidence Trifid members based on Gaia EDR3 parallaxes and proper motions. We also re-analyzed membership of Trifid Main and separated out multiple background stellar associations. Trifid North represents a stellar population $\sim$10\% as rich as Trifid Main that formed in a separate part of the cloud. The 1D stellar velocity dispersion in Trifid North ($0.6\pm0.2$~km~s$^{-1}$) is three times lower than in Trifid Main ($1.9\pm0.2$~km~s$^{-1}$). Furthermore, in Trifid Main, proper motions indicate that the portion of the star cluster superimposed on the optical nebula is expanding. Expansion of the H\,{\sc ii} region around the O-star HD 164492A, and the resulting gas expulsion, can explain both the motions of the stars and gas in Trifid Main. Contrary to previous studies, we find no evidence that a cloud-cloud collision triggered star formation in the region.
\end{abstract}


\section{Introduction}\label{sec:introduction}

In the vicinity of the Sun, many of the major star-forming regions are associated with long, filamentary, molecular-cloud structures \citep[e.g.,][]{Jackson2010,Goodman2014,Zucker2015,Green2019,Lallement2019,Lallement2022,Zucker2020,Alves2020}. One of these filaments is a kpc-long collection of molecular clouds and star-forming regions, traditionally considered to be part of the Milky Way's Sagittarius Arm, but shown by Gaia to form a discrete substructure with a high-pitch angle relative to the arm \citep{Kuhn2021_Sgr}. This structure contains several of the nearest massive star-forming regions, including Messier 8 (the Lagoon Nebula), 16 (the Eagle Nebula), 17 (the Omega Nebula), and 20 (the Trifid Nebula). Numerical galaxy simulations find that filamentary chains of molecular clouds in spiral arms undergo continuous accretion and dispersal \citep[e.g.,][]{2013MNRAS.432..653D,Ibanez-Mejia2017,Lingard2021}. Thus, collisions between clouds may be an important ingredient in star formation in such environments.

\begin{figure*}[t]
\centering
\includegraphics[angle=0.,width=\textwidth]{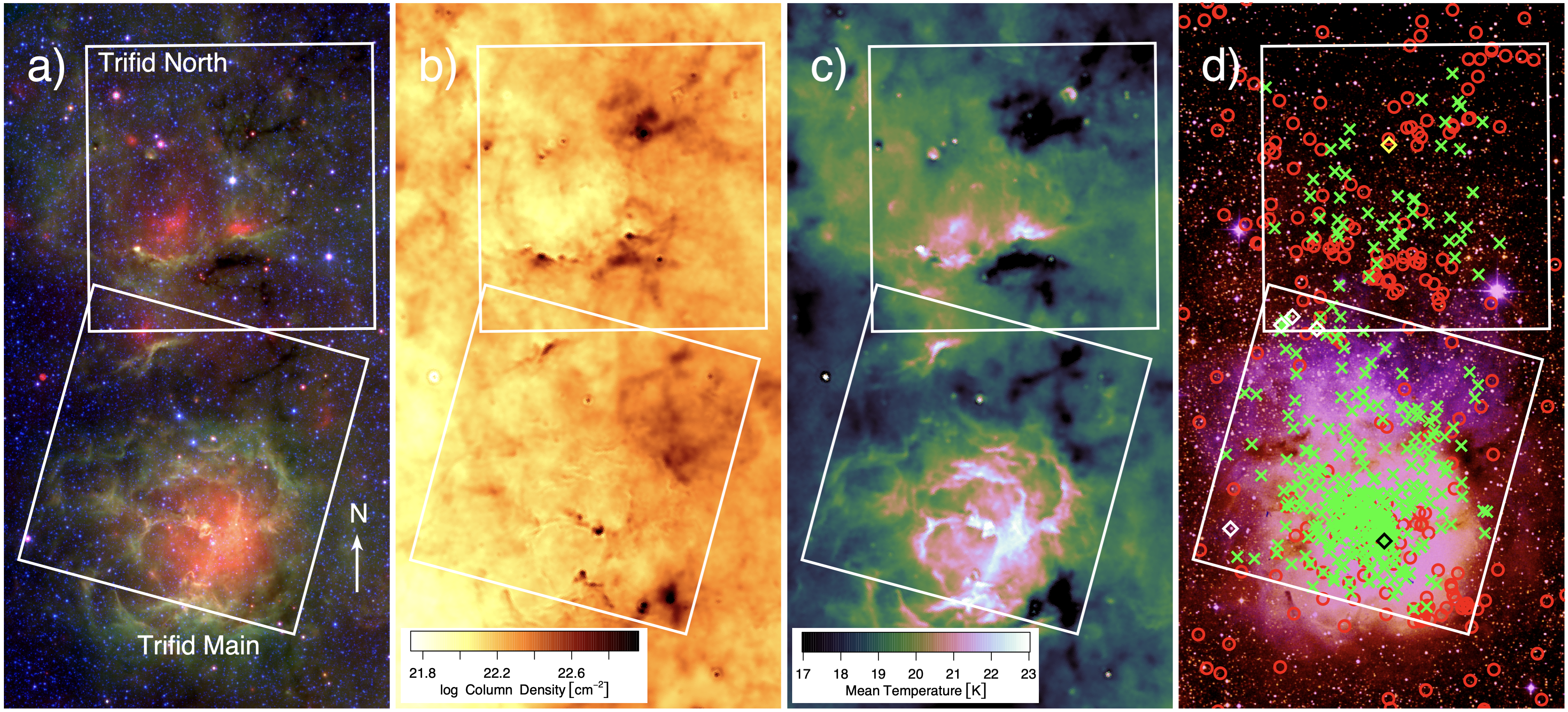}
\caption{ 
Maps of the Trifid region from various observations, showing a rectangular region $\sim0.4^\circ\times0.7^\circ$ centered on $(\alpha,\delta)\sim(270.6^\circ,-22.8^\circ)$, where north is up. The white squares are the Chandra/ACIS-I fields of view. 
a) Mid-infrared Spitzer image generated from 24~$\mu$m (red), 8~$\mu$m (green), and 3.6~$\mu$m (blue) MIPSGAL and GLIMPSE mosaics.
b) Dust column density from the Hi-GAL survey converted to units of $N(\mathrm{H}_2)$ from \citet{Marsh2015}. 
c) Corresponding dust temperatures from \citet{Marsh2015}. 
d) DSS red and blue-band images. X-ray sources identified as cluster members (green x's),  Spitzer/IRAC candidate YSOs (red circles), and massive stars (HD~164492A: black diamond; candidate OB stars: white diamonds; MN68: yellow diamond) are overlaid.
\label{fig:4panel}}
\end{figure*}

What role (if any) cloud-cloud collisions play in star formation is currently an open question \citep[][]{Scoville1986,Fukui2021_sim}. Molecular-line position--velocity (P--V) maps reveal numerous cases where clouds with different velocities appear to be interacting -- often coincident with sites of high-mass star formation -- and this has been interpreted as a sign of cloud-cloud collisions by many studies \citep[][and references therein]{Fukui2021_trigger}. One example is the Trifid Nebula, proposed by \citet{Torii2011,Torii2017} as a case of a cloud-cloud collision triggering star formation.

Here, we examine the young stars in different parts of the Trifid molecular cloud complex to test how the cloud kinematics may have affected star formation across the region. 


\begin{figure}[h]
\centering
\includegraphics[angle=0.,width=0.48\textwidth]{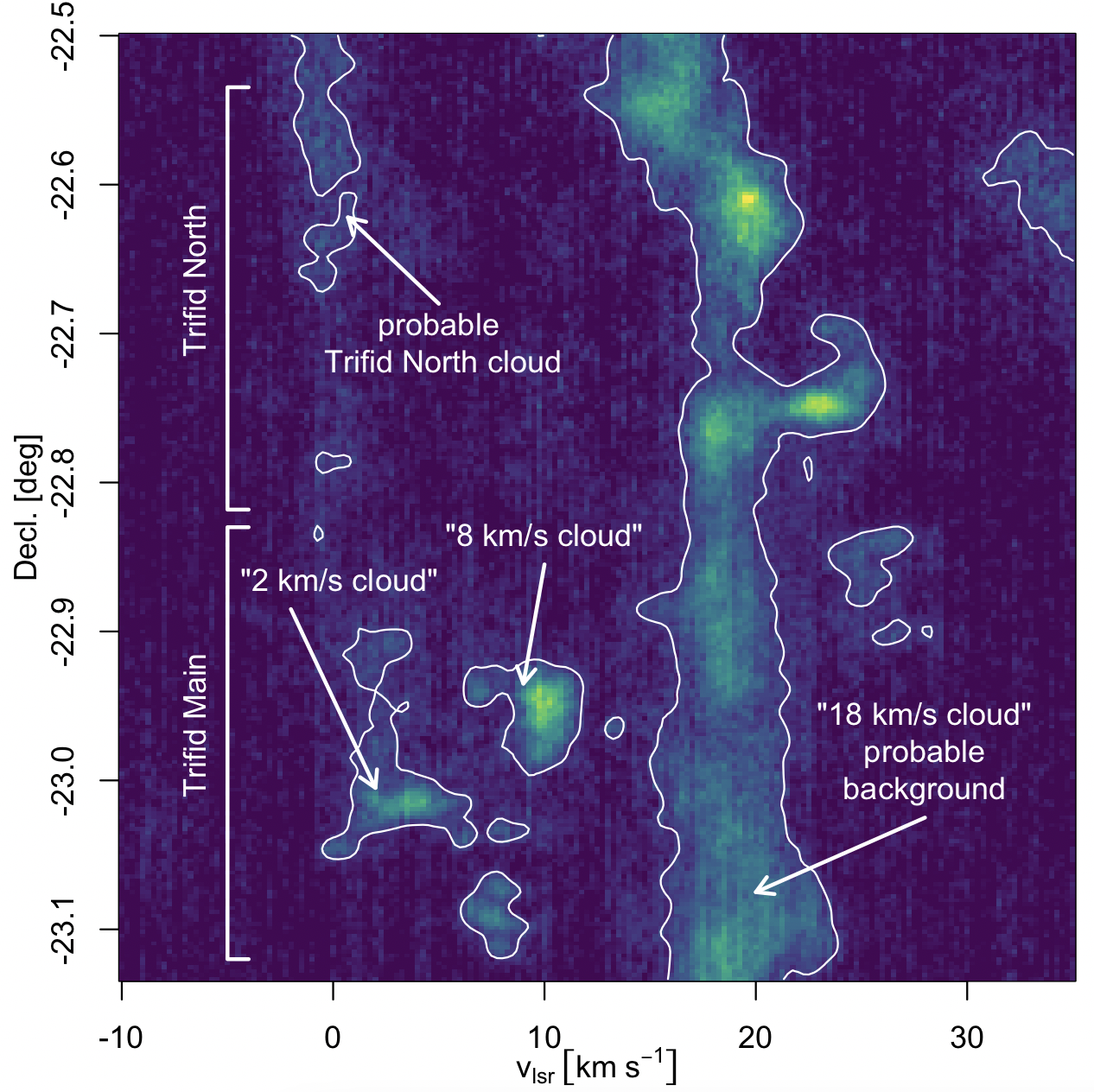}
\caption{$^{13}$CO $J=2$--1 position--velocity (PV) diagram from the SEDIGISM survey \citep{Schuller2021}. The PV slice is centered at $\mathrm{R.A.}=270.5833^\circ$, has a width of 0.25$^\circ$, and spans a range of declinations covering both Trifid Main and Trifid North. The cloud components associated with the Trifid Nebula by \citet{Torii2011} are the ``2 km/s cloud'' and the ``8 km/s cloud.'' The probable extension of this cloud complex to the north is labeled ``Trifid North cloud.'' The positioning of this slice relative to the molecular cloud is shown in Appendix~\ref{sec:cloud}.
\label{fig:sedigism}}
\end{figure}

The Trifid Nebula is an optical- and radio-bright H\,{\sc ii} region, $\sim$30$^\prime$ in diameter, ionized by the late O-star HD~164492A (Figure~\ref{fig:4panel}). The associated young star cluster is trifurcated by dark dust lanes \citep{1833RSPT..123..359H,Rho2008}. Millimeter-wavelength molecular-line observations toward the nebula reveal several cloud components at velocities of $v_\mathrm{lsr}\sim2$~km~s$^{-1}$ (corresponding to the dark lanes crossing the optical nebula), $v_\mathrm{lsr}\sim8$~km~s$^{-1}$ (corresponding to dark features at the edge of the optical nebula), and $v_\mathrm{lsr}\sim18$~km~s$^{-1}$ \citep{Cernicharo1998,Torii2011,Torii2017}. A bridging feature between the 2~km~s$^{-1}$ and 8~km~s$^{-1}$ clouds is coincident with the optical nebula ($\mathrm{decl.}\sim-23^\circ$) and indicates that these clouds are physically interacting and may have collided \citep{Haworth2015,Torii2017,Kalari2021}. However, the cloud's kinematics could alternatively be interpreted as either an expanding H\,{\sc ii} region or turbulence  \citep{2000ApJ...533..911F,Rho2008}. 
A 0~km~s$^{-1}$ cloud continues northward beyond the site of the possible cloud-cloud collision up to $\mathrm{decl.}\sim-22.5^\circ$ (Figure~\ref{fig:sedigism}). The 18~km~s$^{-1}$ cloud is much larger but may be in the background (Appendix~\ref{sec:cloud}).


Since the north region (hereafter Trifid North) is outside the area of the proposed cloud-cloud collision (hereafter Trifid Main), study of both regions allows us to investigate whether the collision (if it occurred) explains any differences in star formation. To identify young stars, we acquired a Chandra X-ray observation of Trifid North, augmenting a previous Chandra observation of Trifid Main by \citet{Rho2004}. X-ray observations have proven effective at selecting members of young embedded star clusters without requiring the stars to have detectable disk or accretion signatures \citep{Feigelson2013}. 

Previous multiwavelength observations of Trifid North show, in projection, indications of star formation  (Figure~\ref{fig:4panel}). An H\,{\sc ii} bubble, [L89b] 7.299-00.116, was detected within the Trifid North field using radio hydrogen recombination lines  \citep{Lockman1989}. There are several infrared-dark clouds, whose rims are bright in Spitzer's 8.0~$\mu$m band, suggesting the presence of an ionizing source to excite emission from polycyclic aromatic hydrocarbons \citep{2006ApJ...643..965R}. Warm dust in Trifid North is seen as 24~$\mu$m emission by Spitzer and is also indicated in Herschel dust-temperature maps \citep{Marsh2015}. An evolved massive star MN68 \citep[][]{Gvaramadze2010,Wachter2010} lies near the center of the Trifid North field, and a small group of candidate OB stars lie between Trifid North and Trifid Main \citep{2017ApJ...838...61P}; see Appendix~\ref{sec:wr} for spectroscopic classification of these sources. Finally, $\sim$90 young stellar objects (YSOs) have been detected in the Trifid North region as part of the Spitzer/IRAC candidate YSO catalog \citep[SPICY;][]{SPICY}. However, the distances to these sources have not previously been established.

Before the Gaia astrometric survey, there had been considerable uncertainty in the distance to the Trifid Nebula, with estimates ranging from $\sim$1.7~kpc \citep{Rho2008} to $\sim$2.7~kpc \citep{Cambresy2011}. Gaia's measurements have dramatically improved the situation, indicating that Trifid is closer than previously thought. We adopt a distance\footnote{The Trifid distance estimate of 1180$\pm$25~pc from \citet{Kuhn2021_Sgr} is based on a joint probabilistic Bayesian model of Gaia EDR3 astrometry for Trifid member stars, adopting the parallax zero-point corrections from \citet{Lindegren2021_zp}. Subsequently, \citet{Kalari2021} published a slightly farther, but less precise, distance estimate of $1257^{+190}_{-98}$~pc, derived by first estimating Bayesian distances to individual stars using a prior similar to that of \citet{2018AJ....156...58B}, then estimating the mean of the distribution of stellar distances. However, their approach effectively uses multiple factors of the same prior (one for each star), which can introduce a strong bias into the estimated cluster distance. \citep[See cautionary note by ][their Section 4.3.]{2021AJ....161..147B} 
} of 1180$\pm$25~pc for Trifid derived by \citet{Kuhn2021_Sgr} using Gaia EDR3 -- this distance is corroborated by our Chandra+Gaia analysis here. Nevertheless, the complicated molecular cloud velocity structure along the line of sight suggests the possibility of the coincidental alignment of multiple star-forming regions with different distances.    

In this paper, Section~\ref{sec:data} describes the data used for the analysis. Section~\ref{sec:mem} provides lists of possible members of the Trifid Complex. Section~\ref{sec:pm_refine} refines membership using Gaia EDR3 proper motions and examines an unrelated background group on the same line of sight. 
Section~\ref{sec:xray} discusses stellar X-ray properties. Section~\ref{sec:oir} discusses optical and infrared properties. Section~\ref{sec:pm} investigates the kinematics of stars in Trifid Main and Trifid North. Section~\ref{sec:discussion} discusses the influence of the molecular cloud properties on stellar kinematics. And, Section~\ref{sec:conclusion} provides the conclusion. 

\section{Data}\label{sec:data}

X-ray emission is ubiquitous from low-mass pre-main-sequence stars and can reveal embedded young stellar populations \citep{Feigelson2018}. We used a combination of X-ray, optical, and infrared source properties to identify candidate young stars in the vicinity of the Trifid Nebula.

\subsection{X-ray Observation and Data Reduction\\ for Trifid North}\label{sec:xray_obs}

The Trifid North field was observed for $\sim$53.8~ks (ObsID 21153; PI M.\ Kuhn) by Chandra on 2019-07-29 using the Advanced CCD Imaging Spectrometer \citep[ACIS-I;][]{2003SPIE.4851...28G} in very faint mode. The aim point of the observation was 18$^h$ 02$^m$ 20.20$^s$ $-$22$^\circ$ 40$^\prime$ 02.00$^{\prime\prime}$ with a roll angle of 269$^\circ$. All four ACIS-I chips were active during the observation, providing a $17^\prime\times17^\prime$ field of view. 

Data reduction was performed using methods developed at Penn State, employing both publicly available software such as ACIS Extract\footnote{\url{http://personal.psu.edu/psb6/TARA/AE.html}} and TARA \citep{Broos2010,2012ascl.soft03001B} and proprietary Level~1 to Level~2 data preparation scripts. These procedures rely on other software packages, including heasoft 6.26 and CALDB 4.9.2.1 \citep{2014ascl.soft08004N}, CIAO 4.12 \citep{2006SPIE.6270E..1VF}, MARX \citep{2013ascl.soft02001W}, AstroLib \citep{1993ASPC...52..246L}, wavdetect \citep{2002ApJS..138..185F}, and SAOImage DS9 \citep{2003ASPC..295..489J}. Versions of this methodology have been used for many X-ray studies, most recently in studies by \citet[][]{2018ApJS..235...43T}, who describe the procedures in their Section 2.2, and \citet{2019ApJS..244...28T}. These methods have been optimized to detect faint X-ray sources (e.g., 3--10 net counts), which dominate the source statistics in observations of star-forming regions, typically detecting $\sim$50\% more X-ray sources than standard reduction recipes from the Chandra X-ray Center.\footnote{\url{https://cxc.cfa.harvard.edu/ciao/threads/}} Extraction apertures were set to include $\sim$90\% of the source flux. Source significance was assessed by the probability, $p_B$, that a source could be a spurious fluctuation in background counts assuming events follow a Poisson distribution; only sources with $p_B<0.01$ are retained. 

The list of 143 X-ray sources is provided in Table~\ref{tab:xray}, with photometric properties for two energy bands, the ``full band'' (0.5--8.0~keV) and the ``hard band'' (2.0--8.0~keV). X-ray fluxes are calculated using Equations~8--9 in \citet{Broos2010}.

X-ray median energy, i.e., the background-corrected median of the event energies for a source within a given band, can serve as an indicator of absorbing column density when calibrated to account for the intrinsic X-ray source spectrum \citep{Getman2010}. Median energy is a detector-dependent quantity, and changes in the quantum efficiency of ACIS due to buildup of contamination on the detector or optical blocking filters\footnote{\url{https://cxc.cfa.harvard.edu/ciao/why/acisqecontamN0014.html}} have diminished ACIS's sensitivity to lower-energy X-ray photons. For pre-main-sequence stars observed by Chandra, this effect has produced a noticeable shift to higher median energies in observations since $\sim$2016. In Table~\ref{tab:mem}, we have subtracted the expected shift (Appendix~\ref{sec:obf}) from the observed median energy to obtain a corrected value that can be directly compared to median energies from previous Chandra observations (e.g., the observation of Trifid Main). 

\subsection{Optical and Infrared Photometry}

We cross-matched the Trifid North X-ray catalog to several optical and infrared catalogs, including Gaia's early third data release \citep[EDR3;][]{GaiaBrown,GaiaCollaboration2021}, the Two Micron All Sky Survey \citep[2MASS;][]{2006AJ....131.1163S}, the United Kingdom Infra-Red Telescope (UKIRT) Infrared Deep Sky Survey \citep[UKIDSS;][]{2007MNRAS.379.1599L,2008MNRAS.391..136L} with photometry from \citet{King2013}, and Spitzer's Galactic Legacy Infrared Mid-Plane Survey Extraordinaire \citep[GLIMPSE;][]{Benjamin2003,Churchwell2009}. Cross-matching is performed by STILTS \citep{2005ASPC..347...29T} using a 1.2$^{\prime\prime}$ match radius. This radius is sufficient to account for uncertainties in both the X-ray positions (0.06$^{\prime\prime}$--1$^{\prime\prime}$; median of $\sim$0.3$^{\prime\prime}$) and in the optical/infrared source positions. The cross-match to Gaia yields 63 matches, whereas, when X-ray positions are artificially shifted 5$^\prime$ west, 7 matches are found, suggesting a spurious-match rate up to $\sim$10\%.\footnote{The mean of the angular separations for matches to the true positions is smaller than for the shifted positions, suggesting that, when both real and spurious counterparts are within the match radius, the real counterpart is likely to take priority.} Spurious matches are unlikely to have the same parallaxes and proper motions as Trifid members, so they are expected to be filtered out at a later stage. 

Gaia provides high-precision astrometry, including parallax ($\varpi$) and proper motion $(\mu_{\alpha^\star},\mu_\delta)$ measurements for just under half the X-ray sources.\footnote{We use the notation $\alpha$ for R.A., $\delta$ for decl., $\mu_{\alpha^\star}\equiv\mu_\alpha\cos{\delta}$ for proper motion in R.A., and $\mu_\delta$ for proper motion in decl.} 
To ensure that Gaia sources have accurate astrometry, we only use astrometric solutions with renormalized unit weight errors $\mathrm{RUWE}<1.4$ and astrometric excess noise $\leq1.0$~mas \citep[see][]{Kuhn2019,Kuhn2020_nap}. We correct parallaxes using the zero-point offsets estimated as a function of source ecliptic latitude, magnitude, and color \citep{Lindegren2021_zp}. These corrections can be important when examining parallax distributions of cluster members in cases where differential absorption produces significant differences in source colors \citep[e.g.,][]{Kuhn2020_rnaas}.

\subsection{Literature YSO Catalogs}

For Trifid Main, we use the list of stars from the Massive Young Star-Forming Complex Study in Infrared and X-ray \citep[MYStIX;][]{Feigelson2013,Kuhn2013,Kuhn2013_ir,King2013,Naylor2013,Povich2013,Broos2013,Townsley2014}. MYStIX used data products similar to those available for Trifid North, including a 58~ks Chandra observation (ObsID 2566), originally published by \citet{Rho2004} and re-analyzed in MYStIX, along with UKIDSS and GLIMPSE catalogs. MYStIX includes 532 probable Trifid members \citep{Broos2013}, which we have cross-matched to Gaia EDR3 using the same procedures as for Trifid North. Since the publication of the MYStIX catalog, \citet{Kalari2021} identified 26 additional H$\alpha$-emission member candidates not in the MYStIX sample.

For both Trifid Main and Trifid North, we also include infrared-excess selected YSOs from the SPICY catalog \citep{SPICY}. The SPICY catalog was generated using a random-forest based strategy to classify mid-infrared sources from the GLIMPSE survey and identify stars with circumstellar disks or envelopes. There are 90 SPICY stars in Trifid North and 105 in Trifid Main. 

\section{Membership Classification and Distance}\label{sec:mem}

In X-ray observations of star-forming regions, contaminants are mainly field stars and extragalactic sources \citep{Getman2011}. 
At this wavelength, field-star counts are significantly reduced compared to the optical because most are not bright enough to be detected. Furthermore, when stars have sufficiently precise Gaia astrometry, objects at different distances or with different proper motions than cluster members can be screened out. Along lines of sight near the Galactic plane, extragalactic sources tend to be highly absorbed and not optically visible, and simulations predict near-infrared magnitudes of $J>20$~mag \citep{Getman2011,Broos2013}. 

\begin{figure}[t]
\centering
\includegraphics[angle=0.,width=0.45\textwidth]{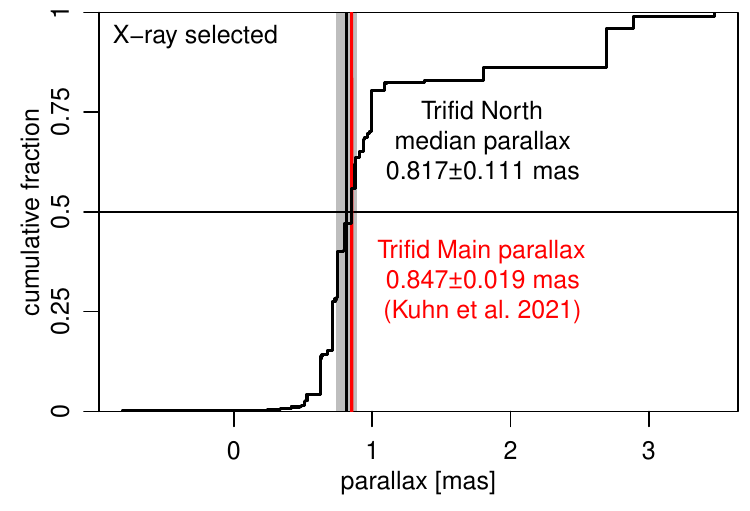}
\caption{Weighted cumulative distribution of parallaxes for Gaia counterparts to X-ray sources in the Trifid North field (black line). Step sizes are proportional to the reciprocal of the parallax error squared. The weighted median parallax and the 1$\sigma$ uncertainty on this quantity are indicated by the vertical black line, and the gray envelope. The median parallax of the X-ray sources in Trifid North is consistent with the parallax of the Trifid Main cluster (red line). 
\label{fig:plxmed}}
\end{figure}

\begin{figure*}[t]
\centering
\includegraphics[angle=0.,width=0.48\textwidth]{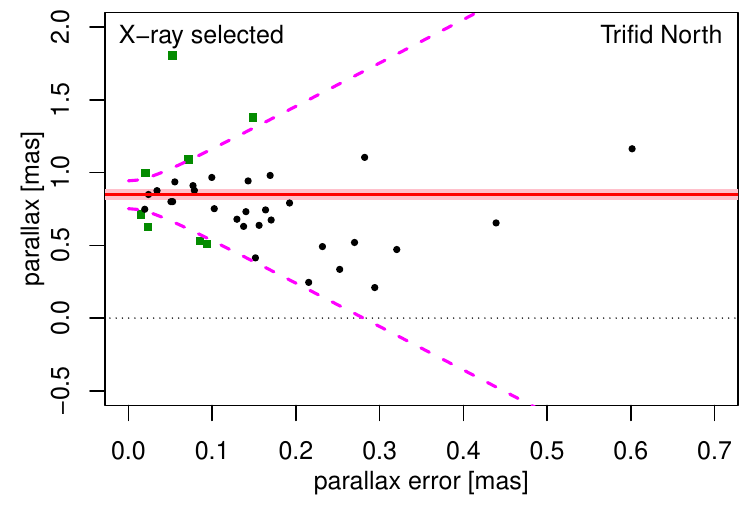}
\includegraphics[angle=0.,width=0.48\textwidth]{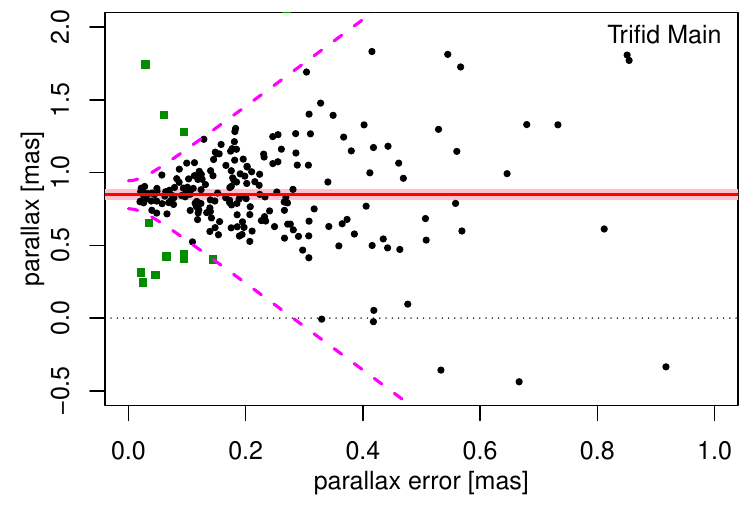}
\caption{Scatter plot of parallax versus parallax error for stars in the Trifid North field (left) and the main cluster in the Trifid Nebula (right). The parallax of the Trifid Nebula is indicated by the horizontal red line. The dashed magenta lines encompass the region of this diagram expected to include 99.7\% ($3\sigma$) of stars at the same distance. Stars outside this region (green squares) are probable nonmembers, while stars inside (black points) are possible members.
\label{fig:plxmem}}
\end{figure*}

\subsection{Trifid North}

We examine the parallax distribution of Gaia counterparts to X-ray sources to investigate whether they are associated with the Trifid Complex. Figure~\ref{fig:plxmed} shows the weighted cumulative distribution of the parallaxes of these sources, where $1/\mathtt{parallax\_error}^2$ is used as a weight. The weighted median is $0.817\pm0.111$~mas, and the cumulative distribution rises sharply around this value, indicating that many of these X-ray sources are located at approximately the same distance. This value is statistically consistent with the parallax of $0.847\pm0.019$~mas found by \citet{Kuhn2021_Sgr} for Trifid Main. The similarity of these values strongly suggests that the X-ray sources in Trifid North are mainly members of the Trifid complex, with perhaps some foreground and some, but relatively little, background contamination. Given that the measurements for Trifid Main are more precise, we adopt $1180\pm25$~pc as the distance for the entire Trifid complex. 

Our membership classifications for X-ray sources in Trifid North include several tiers, depending on the evidence available. To be considered a possible member, an X-ray source source must be either detected by Gaia or have a UKIDSS $J$-band magnitude $<20$~mag. Furthermore, if a source has good Gaia astrometry, its parallax must be consistent with the distance to the Trifid Nebula (Figure~\ref{fig:plxmem}, left panel). Sources with small parallax uncertainties are required to be closer to the Trifid value than those with larger uncertainties. We reject the membership of any source with a parallax discrepant by more than 3 standard deviations from our adopted Trifid Nebula value, taking into account  measurement errors, the uncertainty in the parallax of Trifid, and the 0.026~mas systematic uncertainty in parallax zero point. The 3$\sigma$ threshold is expected to encompass 99.7\% of sources at the distance of Trifid, so it is extremely unlikely that any object outside this region could be a member. Our list of 51 candidate members is given in Table~\ref{tab:mem}.

Intriguingly, most of the rejected sources (7 out of 8) lie very close to the 3$\sigma$ boundary. Given that the Trifid Nebula is part of a larger kpc-scale star-forming structure in the Sagittarius Arm \citep{Kuhn2021_Sgr}, it is possible that some of these stars may be part of this larger scale structure, albeit not part of Trifid. 

\subsection{Trifid Main}

For Trifid Main, we use the same strategy as above to refine the published MYStIX membership list (Figure~\ref{fig:plxmem}, right panel). The objects with the smallest parallax errors are clustered symmetrically around the estimated $\varpi=0.847$~mas Trifid parallax from \citet{Kuhn2021_Sgr}, supporting this previous result. Out of 532 sources, 17 ($\sim$3\%) are rejected as nonmembers based on parallax. The refined Trifid membership list is given in Table~\ref{tab:trifid}. The 26 new H$\alpha$ sources from \citet{Kalari2021} are also consistent with our adopted parallax.

\subsection{Infrared-Excess Sources in Trifid North}\label{sec:irexcess_plx}

While the X-ray selected stars appear to be mostly at the distance of the Trifid Nebula, this is not true for the infrared-excess selected SPICY stars in Trifid North. Out of 90 SPICY stars in Trifid North, only 8 have good Gaia astrometry, and the weighted mean parallax for these objects is 0.37~mas, implying that they are more distant. We used the weighted $t$-test \citep{weights2016} to determine whether this parallax inconsistency is statistically significant, finding that both the difference in means of SPICY versus Trifid Main ($p<10^{-5}$) and SPICY versus the Trifid North X-ray sample ($p<10^{-5}$) are highly significant. 

A parallax of 0.37 happens to match the pre-Gaia 2.7~kpc distance estimate by \citet{Cambresy2011}, hinting that \citeauthor{Cambresy2011}'s method may have been sensitive to the background star-forming region rather than to the Trifid Complex. 

Thus, in the Trifid North field, X-ray and infrared-excess selection appear to be sensitive to two different populations at two different distances. (There are only 3 sources in common, none of which have good Gaia astrometry.) This situation could arise if the pre-main-sequence stars in Trifid North are slightly older, and have lost their disks, so most of the disk-bearing stars are in the background. Meanwhile, the X-ray sources tend to be faint, so Chandra is largely insensitive to pre-main-sequence stars at $\sim$2.7~kpc. For these reasons, we focus the analysis on only the X-ray selected member candidates.  

\begin{figure*}[t]
\centering
\includegraphics[angle=0.,width=0.7\textwidth]{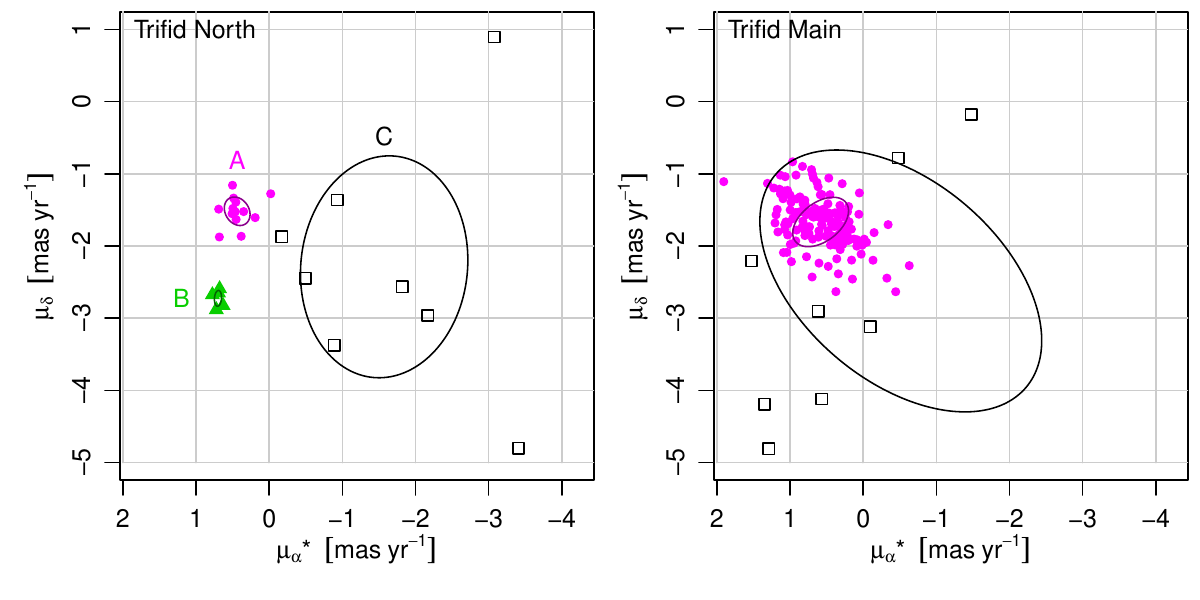}
\caption{Scatter plots showing proper motions of probable members in Trifid North (left) and Trifid Main (right). Only sources with good Gaia astrometry (see text) are shown. The groups identified by the Gaussian mixture model are indicated by the ellipses (Mahalanobis distance of 1) and the color-coded points. The kinematic groups in Trifid North include: Group A (13 stars) having proper motions similar to Trifid Main, Group B (4 stars) having discrepant proper motions, and Group C (8 stars) whose proper motions resemble field stars.  
\label{fig:pm_mem}}
\end{figure*}

\section{Refining Membership with Proper Motions}\label{sec:pm_refine}

Stellar siblings are expected to be born with similar Galactic orbits, so the clustering of stars in proper-motion space can be used to refine membership. This strategy helps generate a pristine sample by removing remaining field stars that happen to be at the same distance as the star-forming region. However, it may cause us to miss high-velocity cluster members, such as run-away or walk-away stars.

The proper motion distributions for X-ray selected members of Trifid Main and Trifid North are shown in Figure~\ref{fig:pm_mem}, using only stars with high-precision astrometry ($\mathtt{astrometric\_sigma5d\_max} \leq 0.5$~mas). In both cases, some stars appear clustered in proper-motion space, while others are more distributed. 

We performed unsupervised cluster analysis on these distributions using Gaussian mixture models implemented by the {\it mclust} software package \citep[][]{fraley2006mclust}. The models are fit using the expectation--maximization algorithm, and minimization of the Bayesian Information Criterion \citep[BIC;][]{schwarz1978estimating} determines the number of components. 

For Trifid Main (Figure~\ref{fig:pm_mem}, right panel), two components are required -- one for stars clustered at $\mu_{\alpha^\star}\sim0.6$~mas~yr$^{-1}$ and $\mu_{\delta}\sim-1.7$~mas~yr$^{-1}$ and a second for stars that are not strongly clustered in proper motion (likely nonmembers). The mixture model calculates classification probabilities for each star, and we assign the stars to the components based on which has the higher probability. From this model, the expected contamination rate for objects assigned to the ``clustered'' component is $\sim$1.5\%. Thus, we can consider the 149 stars belonging to this component to be high-confidence members of Trifid Main.
  
For Trifid North (Figure~\ref{fig:pm_mem}, left panel), three components are needed to fit the proper motion distribution:
\begin{description}
\vspace*{-0.2cm}
\item[Group A] A group of 13 stars with a mean proper motion $(\mu_{\alpha^\star}, \mu_{\delta}) = (0.4,-1.5)$~mas~yr$^{-1}$. 
\vspace*{-0.2cm} \item[Group B] A tight group of 5 stars with $(\mu_{\alpha^\star}, \mu_{\delta}) = (0.7,-2.7)$~mas~yr$^{-1}$.
\vspace*{-0.2cm} \item[Group C] The remaining stars that do not exhibit proper-motion clustering.
\end{description}

\vspace*{-0.2cm} 
The proper motion of Group~A is nearly identical to Trifid Main, so we interpret it as representing the stars associated with the Trifid complex. The model estimates a contamination rate for Group~A of $\sim$1\%, so we can regard these stars as high-confidence members. On the other hand, the stars making up Group~C are more likely to be field stars. Although Group~B only includes 5 stars, the tight clustering of these objects make the existence of this group statistically significant. The difference in BIC for a 3 vs.\ 2-component model is $\Delta$BIC=20, which indicates that the 3-component model is highly favored. 

\begin{figure}[h]
\centering
\includegraphics[angle=0.,width=0.45\textwidth]{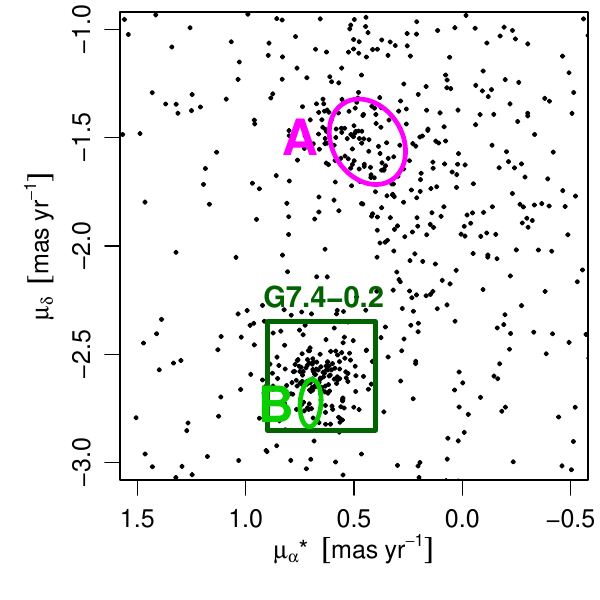}
\caption{Proper motions of Gaia sources (black points) within a $\sim$0.25 square degree region around Trifid North. Only sources with $0.64\leq \varpi \leq 0.92$~mas and meeting the Gaia quality criteria are displayed. For comparison with Figure~\ref{fig:pm_mem}, the ellipses from that figure have been overlaid. Density enhancements in the proper-motion distribution of Gaia sources can be seen near both groups A and B. 
\label{fig:pm_field}}
\end{figure}

\begin{figure}[t]
\centering
\includegraphics[angle=0.,width=0.45\textwidth]{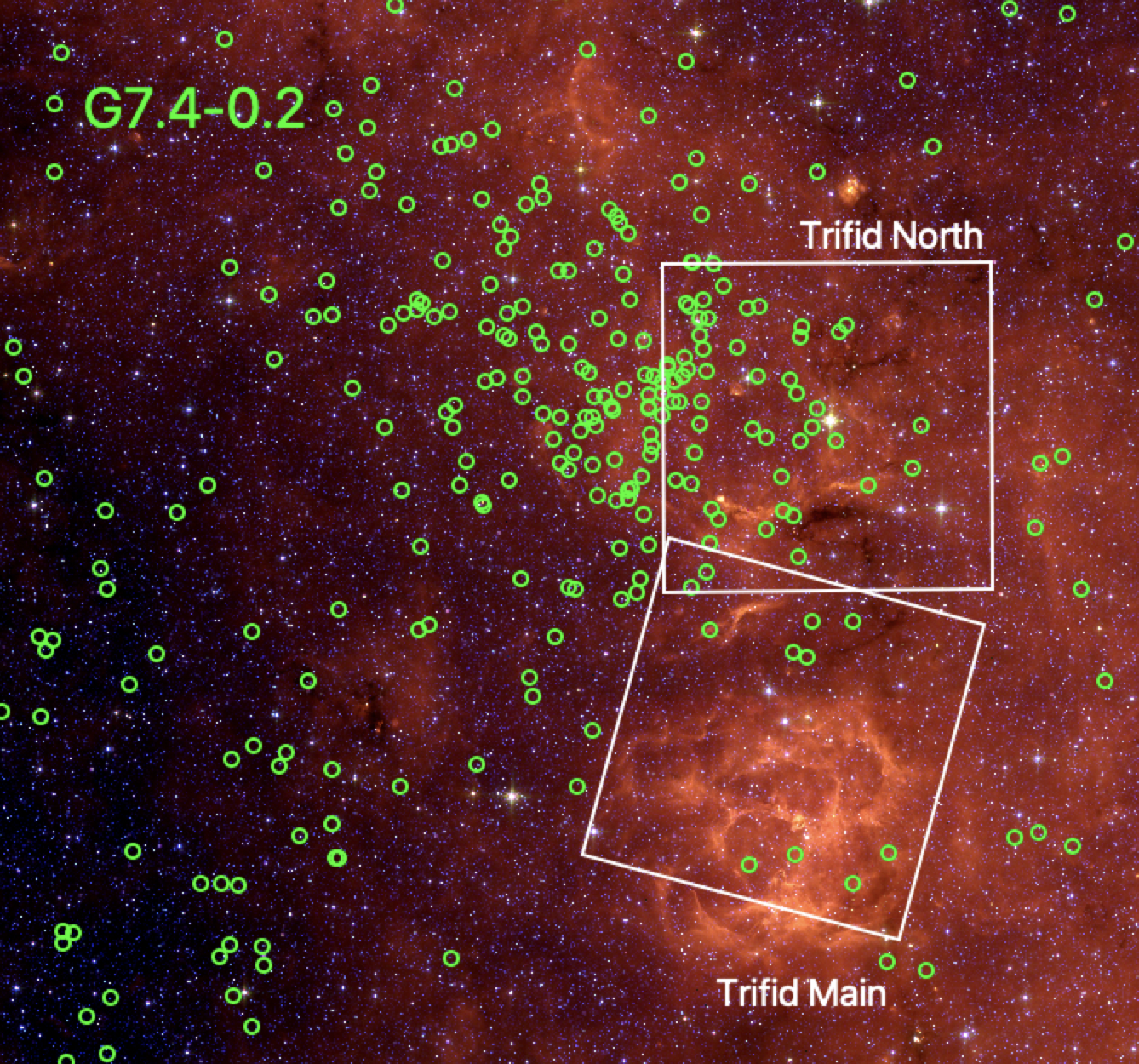}
\caption{Members of G7.4-0.2, selected using the green box in Figure~\ref{fig:pm_field}, are marked by green circles. The background image is from GLIMPSE (red: 8.0~$\mu$m, green: 5.8~$\mu$m, blue: 3.6~$\mu$m). Parallax estimates for this group suggests that it is $\sim$130~pc farther than the stars in the Trifid and Trifid North regions.
\label{fig:groupB}}
\end{figure}

\begin{figure}[h]
\centering
\includegraphics[angle=0.,width=0.45\textwidth]{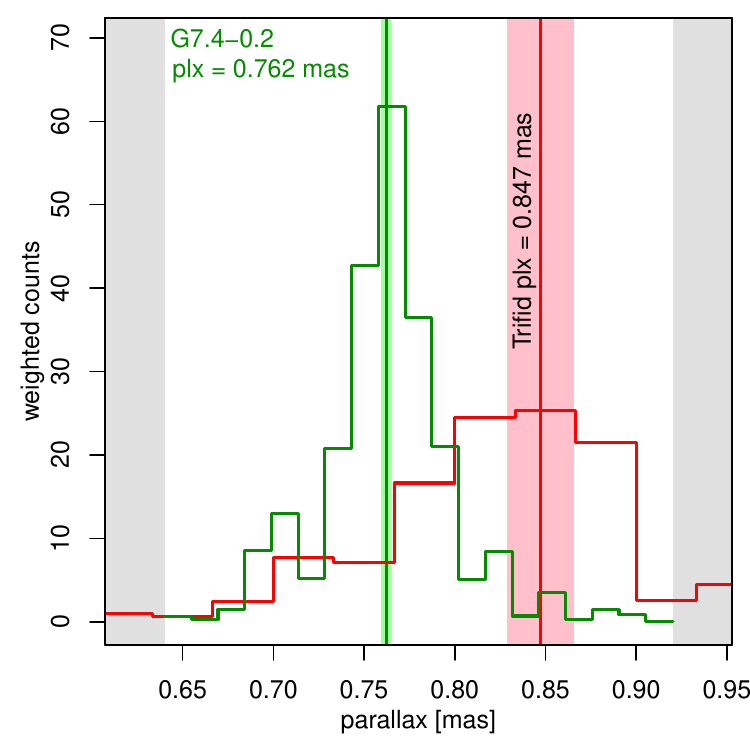}
\caption{Weighted histogram (green) of parallaxes for objects in G7.4-0.2 (corresponding to Group~B), with the vertical line indicating the weighted median with 1$\sigma$ uncertainties. For comparison, the parallax of Trifid Main and a histogram of high-confidence Trifid members are shown in red. The shaded gray regions are the limits used when selecting G7.4-0.2 members. 
\label{fig:groupB_dist}}
\end{figure}

\subsection{A Background Stellar Group}

Group~B can be explained by an unrelated stellar association, designated here G7.4-0.2, which lies a relatively small distance behind the Trifid star-forming region. 

Figure~\ref{fig:pm_field} shows the proper motions of field stars within the neighborhood of Trifid North; only Gaia sources with $0.64~\mathrm{mas} \leq \varpi \leq 0.92~\mathrm{mas}$ and meeting the same astrometric quality criteria as earlier are shown. For reference, the ellipses marking the Groups A and B from Trifid North are overlaid. There is a density enhancement near Group A, but the most prominent feature is a distinct density enhancement near Group B. 

To investigate this density enhancement, we select the Gaia sources with proper motions $0.4 \leq \mu_{\alpha^\star} \leq 0.9$~mas~yr$^{-1}$ and $-2.85 \leq \mu_{\delta} \leq -2.35$~mas~yr$^{-1}$ (the green box in Figure~\ref{fig:pm_field}). The selected sources are spatially clustered, overlapping the eastern edge of the Trifid North field and extending to the northeast (Figure~\ref{fig:groupB}). The distinct density enhancement in proper motion space combined with their spatial clustering suggests that they represent a stellar moving group, which we designate G7.4-0.2. Given that the 5 X-ray selected stars in Group~B have proper motions consistent with this group, but discrepant from Trifid Main stars and the majority of the Trifid North stars, it seems likely that they belong to G7.4-0.2. The 232 Gaia members of G7.4-0.2 are listed in Table~\ref{tab:groupB}. 

The peak of the parallax distribution for G7.4-0.2 is distinctly lower than the parallax of the Trifid cluster (Figure~\ref{fig:groupB_dist}). The weighted median parallax for G7.4-0.2 is $0.762\pm0.003$~mas, corresponding to 1310~pc, which places this association $\sim$130~pc behind the Trifid Nebula. 

\begin{figure*}[t]
\centering
\includegraphics[angle=0.,width=0.48\textwidth]{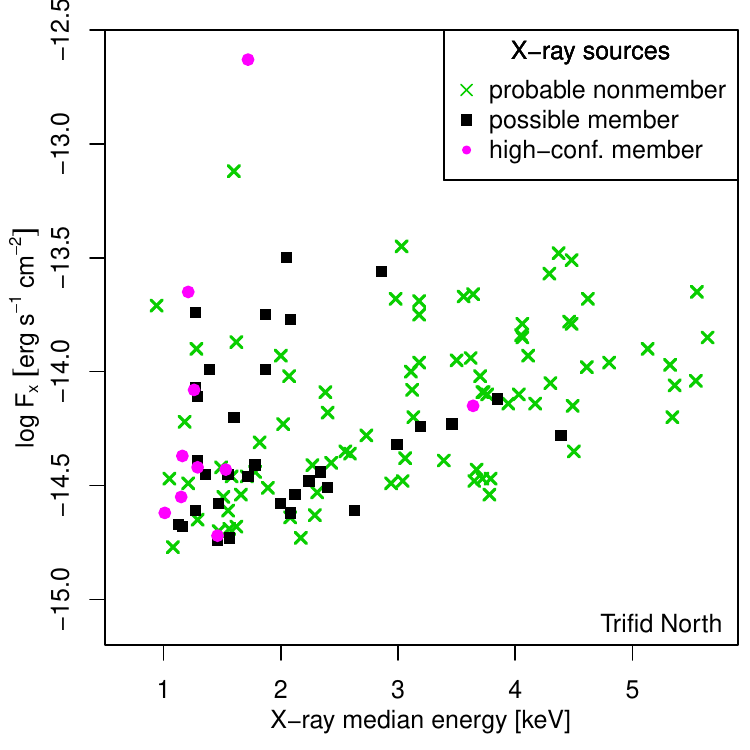}
\includegraphics[angle=0.,width=0.48\textwidth]{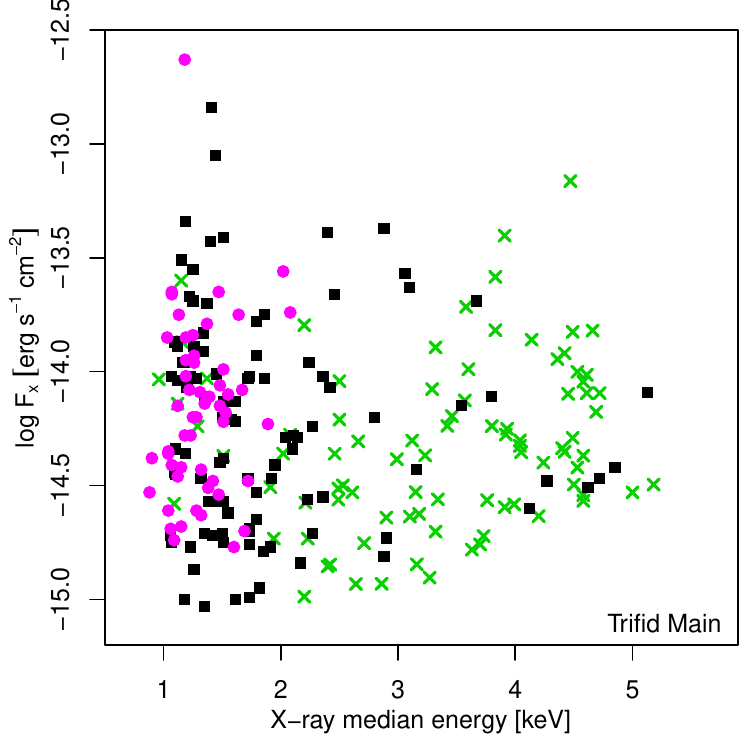}
\caption{X-ray color--magnitude diagrams for Trifid North (left) and Trifid Main (right). X-ray flux in the 0.5-8.0~keV band ($F_X$) is plotted against corrected X-ray median energy. Probable nonmembers are color-coded green, possible members are black, and high-confidence member candidates are magenta. 
\label{fig:xcmd}}
\end{figure*}

\begin{figure}[h]
\centering
\includegraphics[angle=0.,width=0.45\textwidth]{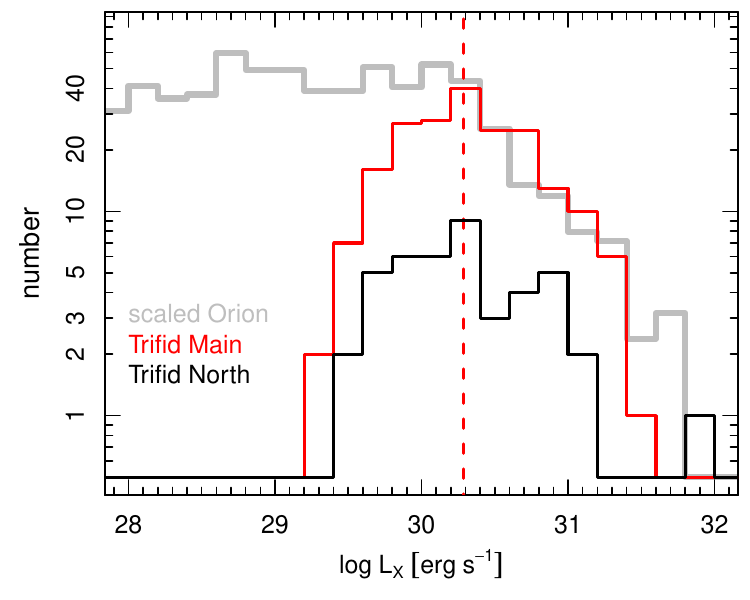}
\caption{ 
Histograms of X-ray luminosities for probable members of Trifid Main (red) and Trifid North (black) using all stars classified as possible members. These are compared to the X-ray luminosity function (XLF) from a sample of pre-main-sequence stars from a deep Chandra observation of the Orion Nebula that is complete down to $L_X\sim10^{28}$~erg~s$^{-1}$. This XLF is scaled (gray line) to match Trifid, from which we estimate a completeness limit of $L_X\sim10^{30.3}$~erg~s$^{-1}$ for both Trifid and Trifid North. 
\label{fig:xlf}}
\end{figure}

\section{X-ray Properties}\label{sec:xray}

X-ray emission from pre-main-sequence stars is generally attributed to stellar coronae heated by magnetic reconnection, and it is typically $10^2$--$10^5$ times higher than emission from main-sequence stars  \citep{Feigelson1999,Gudel2004,Preibisch2005}. 

The X-ray color--magnitude diagrams (i.e., flux vs.\ median energy) are shown in Figure~\ref{fig:xcmd} for both the Trifid North and Trifid Main. X-ray fluxes in Trifid North range from $10^{-14.75}$~erg~s$^{-1}$~cm$^{-2}$ to $10^{-12.5}$~erg~s$^{-1}$~cm$^{-2}$, and median energies range from $\sim$1~keV to $\sim$5~keV, both similar to what is found in Trifid Main. On these diagrams, interstellar absorption shifts sources to lower fluxes and higher median energies. In both regions, most sources classified as member candidates have median energies $<$2~keV; this is similar to the range 1--1.75~keV for intrinsic median energies of pre-main-sequence stars \citep{Getman2010,Kuhn2017}. However, in both cases, a handful of member candidates have higher median energies. These results suggest that in both Trifid Main and Trifid North have high differential absorption. 

X-ray luminosities, $L_X$, and absorption column densities, $N_H$, are estimated with XPHOT \citep{Getman2010}, which uses empirical relations for X-ray fluxes and median energies of pre-main-sequence stars to calculate absorption-corrected luminosities, taking into account both statistical and systematic errors. Applying this method to Trifid North, X-ray luminosities range from $L_X = 4\times10^{29}$~erg~s$^{-1}$ to $7\times10^{31}$~erg~s$^{-1}$ and absorbing columns range from $N_H = 1.4\times10^{20}$~cm$^{-2}$ to $1.6\times10^{23}$~cm$^{-2}$. We also recalculated the X-ray luminosities for the X-ray sources in Trifid Main \citep{Broos2013}, using the updated Gaia-based distance of 1180~pc (Table~\ref{tab:trifid}). 

Figure~\ref{fig:xlf} shows histograms of $L_X$ for candidate members in both Trifid Main and Trifid North, along with a scaled X-ray luminosity function derived from a sample of $\sim$800 lightly absorbed Orion pre-main-sequence stars that is approximately complete down to $L_X\sim10^{28}$~erg~s$^{-1}$ \citep{Getman2005}. The gray line has been shifted vertically on this log-log plot to match the high-luminosity end of the Trifid Main luminosity function. Examinations of X-ray luminosity functions in multiple star-forming regions have found that most have similar shapes at the high-luminosity end \citep[][and references therein]{Kuhn2015}. Furthermore, median X-ray luminosities of pre-main-sequence stars stay approximately constant during the first 3~Myr \citep{Getman2022}. Thus, the X-ray completeness limits for the Trifid samples can be estimated by comparing their empirical $L_X$ distributions to the Orion sample to identify the luminosity below which they have relatively fewer stars \citep[][]{Kuhn2017}. From this strategy, both Trifid North and Trifid Main have completeness limits of $2\times10^{30}$~erg~s$^{-1}$. It is not surprising that both X-ray observations have similar sensitivity, given that both have similar integration times and similar ranges of absorptions.

If we assume that the X-ray luminosity function for the Trifid Nebula is not too dissimilar from the Orion Nebula Cluster, we can scale from the Orion sample to extrapolate the number of missing stars in Trifid. This yields a estimate of $\sim$670 stars in Trifid Main. Given that the histograms in Figure~\ref{fig:xlf} are constructed from possible members, rather than the much smaller samples of Gaia-validated high-confidence members, they should be regarded as upper limits. 

There are 13 Gaia high-confidence Trifid North members versus 149 high-confidence Trifid Main members, both of which were selected in nearly identical ways, and should therefore have been subject to similar selection effects. We can roughly estimate the ratio of total number of stars in Trifid North to Trifid Main as the ratio of these samples $13 / 149 \approx 9\%$. Given our previous estimate for the population of  Trifid Main, this would correspond to a total of $\sim$60 stars in Trifid North. 

\begin{figure*}[t]
\centering
\includegraphics[angle=0.,width=0.75\textwidth]{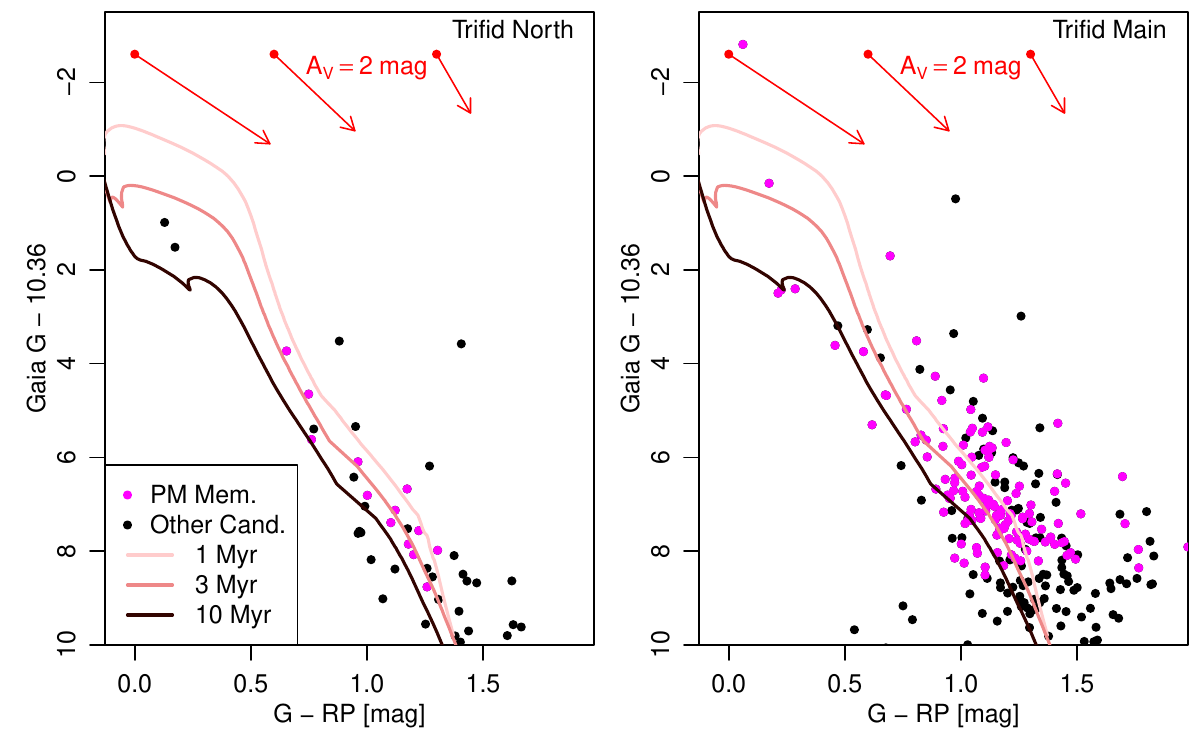}
\caption{ 
Gaia color--magnitude diagram for candidate members in Trifid North (left) and Trifid Main (right). The high-confidence members clustered in proper motion space are magenta points, and other possible members are black. 
The distance modulus of 10.36 has been subtracted from the $G$-band magnitude, but no correction for extinction as been applied. Red arrows show approximate extinction vectors based on the \citet{Cardelli1989} extinction law for sources with intrinsic colors of $G-RP=0$, 0.6, and 1.3~mag. We also include the \citet{Bressan2012} isochrones for 1~Myr, 3~Myr, and 10~Myr for comparison.
\label{fig:gaia_cmd}}
\end{figure*}

\begin{figure*}[t]
\centering
\includegraphics[angle=0.,width=0.7\textwidth]{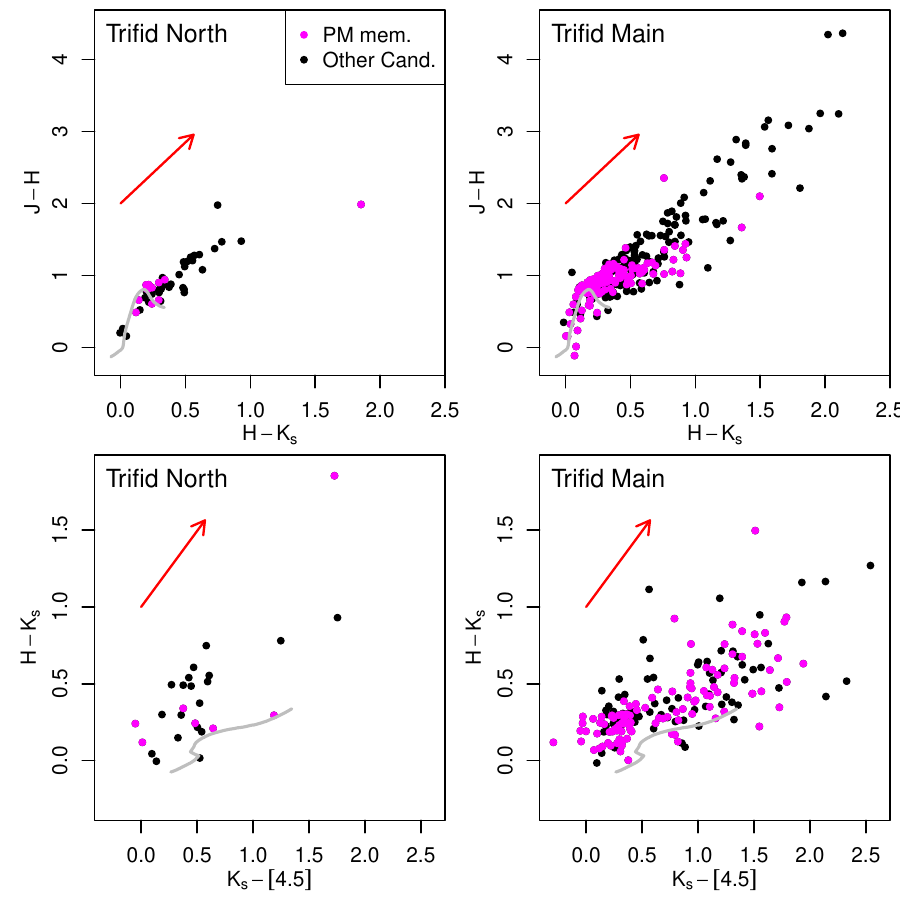}
\caption{Infrared color-color diagrams indicating member candidates selected via their X-ray emission for Trifid North (left) and Trifid Main (right). High confidence proper-motion members are color-coded magenta, while other member possible members are black. The arrows show reddening vectors  corresponding to $A_K = 1$~mag are indicated by the arrows \citep{Rieke1985,Indebetouw2005}. Isochrones for an age of 1~Myr \citep{Bressan2012} are indicated by the gray curves.
\label{fig:ir}}
\end{figure*}

\section{Optical and Infrared Properties}\label{sec:oir}

Figure~\ref{fig:gaia_cmd} shows the distance-corrected Gaia $G$ vs.\ $G-RP$ color-magnitude diagrams for the X-ray selected member candidates in both Trifid North and Trifid Main. Pre-main-sequence stars are expected to lie above the main sequence, and some Gaia-based studies have used this criterion to select candidate young stars \citep[e.g.,][]{Zari2018,Kerr2021,McBride2021}. For all member candidates, we assume a distance modulus of 10.36, since not all stars have parallax measurements, and compare their locations on this diagram to pre-main-sequence isochrones from the  PARSEC models \citep{Bressan2012}. The reddening vectors (which vary as a function of intrinsic source color) are also indicated on Figure~\ref{fig:gaia_cmd}. 
 
In both Trifid Main and Trifid North, most objects lie above the 10~Myr isochrone. In Trifid North, the high-confidence members lie nearest the 3~Myr isochrone, while the Trifid Main stars are more scattered. These diagrams have not been corrected for extinction, which can shift Gaia photometry such that stars appear younger.\footnote{ 
We investigated whether additional photometry from public surveys can provide improved inference of stellar parameters by fitting the spectral energy of Trifid members using the VOSA software \citep{Bayo2008}. Given the large ranges of extinctions in Trifid Main and Trifid North, we found that it is difficult, with only broad-band photometry, to break the degeneracy between extinctions, effective temperatures, and stellar luminosities.
}
Thus, $\sim$3~Myr can be considered a lower limit on the median age of the high-confidence Trifid North stars. This is slightly older than the typical literature age estimates for stars in Trifid Main, such as a mean isochronal age of $\sim$1.5~Myr for pre-main-sequence stars \citep{Kalari2021} and a spectroscopic age of 0.6~Myr for the O star HD~164492A \citep{Petit2019}.

Figure~\ref{fig:ir} shows infrared color-color diagrams for member candidates in Trifid North and Trifid Main -- we only include the X-ray selected sources so that samples for both regions will be comparable. Stars on the $J-H$ vs.\ $H-K_s$ diagrams are spread out in a direction parallel to the extinction vector. These diagrams suggest that most members have extinctions in the range $A_K\sim0$--1~mag, but there is a significant tail of the distribution extending to $A_K\sim3$~mag among the Trifid Main stars. The high-confidence members tend to have lower extinctions, since they must be visible to Gaia. 

Only one high-confidence Trifid North member shows evidence of excess in the near-infrared (Figure~\ref{fig:ir}). This star, 2MASS~J18020358-2235401, has $H-K \approx 1.9$, which is too red to be expained by extinction of a stellar photosphere using the \citet{Bressan2012} stellar models and the \citet{Rieke1985} reddening vector, suggesting that circumstellar material contributes to the $K$-band emission. The half-life for circumstellar disks observed in this wavelength range has been estimated to be $\sim$1--3.5~Myr \citep{Mamajek2009,Ribas2014,Richert2018}, so the low disk fraction in Trifid North is consistant with an age older than $\sim$3~Myr.

\section{Stellar Kinematics}\label{sec:pm}

\begin{figure}[t]
\centering
\includegraphics[angle=0.,width=0.45\textwidth]{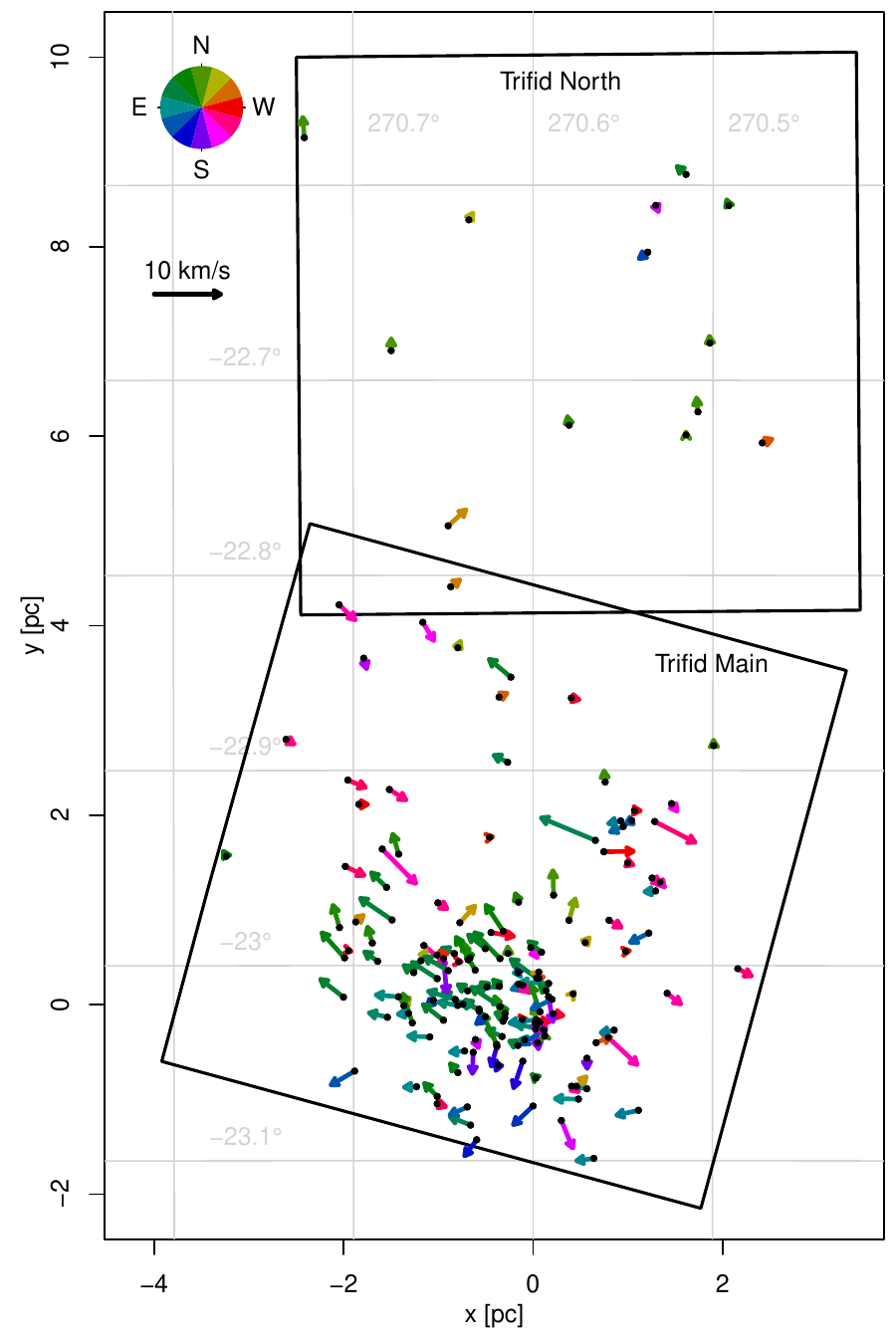}
\caption{Positions of high-confidence members of Trifid Main and Trifid North projected in the Cartesian coordinate system defined in Section~\ref{sec:pm}. Arrows are proportional to tangential velocity, and color-coded based on the direction they are pointing. The mean motion of the Trifid Main cluster is adopted as the rest frame. 
\label{fig:arrows}}
\end{figure}

To convert stellar proper motions into tangential velocities in the Trifid reference frame, we use the strategy from \citet[][their Equations 1--4]{Kuhn2019}, to project the stars onto a Cartesian coordinate system and remove any perspective expansion/contraction induced by the radial velocity of the system. For these corrections, we assume the the radial velocity of the $v_\mathrm{lsr} \approx 2$~km~s$^{-1}$ cloud, which we convert to a heliocentric velocity. The perspective corrections are small enough that some uncertainty in the mean radial velocity of the system does not change the main results. In the resulting Cartesian coordinate system, the $x$ and $y$ axes are parallel to lines of constant R.A.\ and decl.\ at the center of the Trifid Nebula, and $v_x$ and $v_y$ are the corresponding velocities in units of km~s$^{-1}$ (Figure~\ref{fig:arrows}). 

Velocity dispersions can be modeled by multivariate Gaussian distributions, using a statistical model that takes into account the measurement errors provided in the Gaia EDR3 table. Using Equation~13 from \citet{Kuhn2019}, we find that high-confidence members of Trifid North have a one-dimensional velocity dispersion of $\sigma_\mathrm{1D} =0.6\pm0.2$~km~s$^{-1}$ and that the high-confidence members of Trifid Main have a velocity dispersion of $\sigma_\mathrm{1D} =1.9\pm0.1$~km~s$^{-1}$. 

These results demonstrate that the velocity dispersion in Trifid North is particularly low -- lower than the velocity dispersion in Trifid Main and lower than any of the velocity dispersions observed in the $\sim$20 clusters and associations examined by \citet{Kuhn2019}. 

\subsection{Expansion of the Trifid Main Cluster}

In massive star-forming regions,
\citet{Kuhn2019} demonstrated that most groups of young stars visible to Gaia are expanding. This result has been corroborated in many subsequent studies \citep[e.g.,][]{Wright2019,Swiggum2021,Kounkel2021}. Trifid Main was included in this original Gaia DR2 based study, but the expansion velocity of $0.33\pm0.37$~km~s$^{-1}$ was not statistically significant. Gaia EDR3 provides more stars with high-precision astrometry, enabling a more detailed look at this cluster.

\begin{figure*}[t]
\centering
\includegraphics[angle=0.,width=1\textwidth]{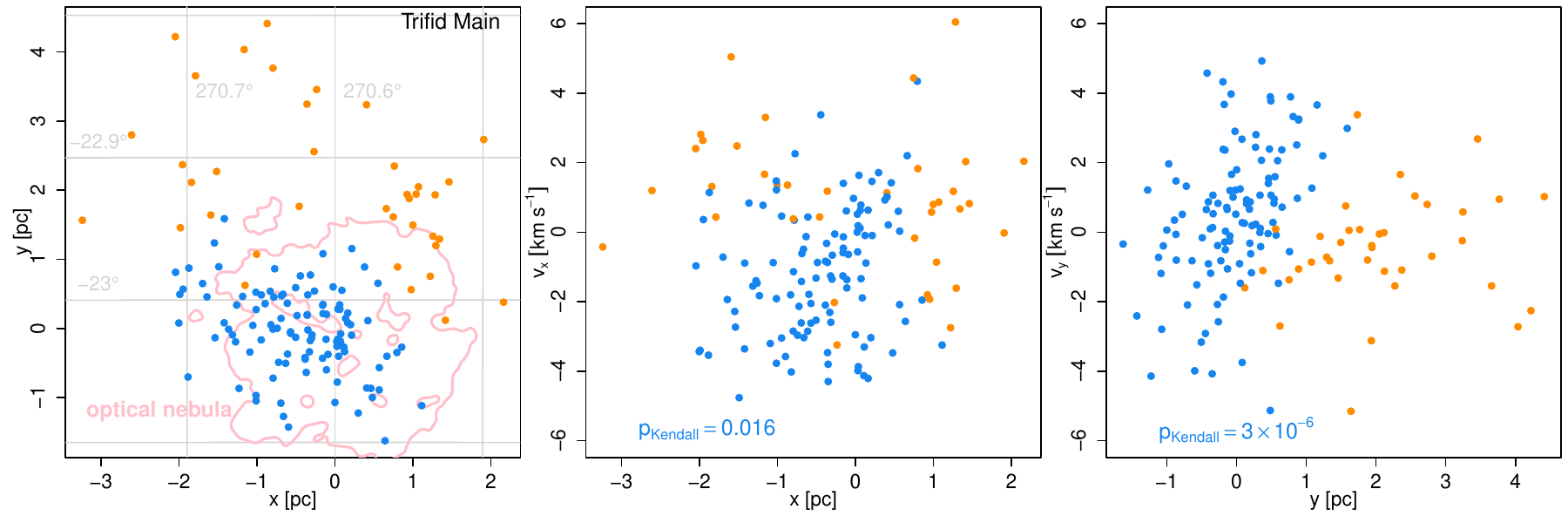}
\caption{Gaia-selected high-confidence members of Trifid Main in position and velocity space. The grouping of stars into two subclusters via a 4-dimensional $(x,y,v_x,v_y)$ Gaussian mixture model is indicated by the color of the points (either orange or blue). Left panel: Spatial distribution of stars. The outline of the optically bright emission nebula is shown in pink. Center and right: Position versus velocity in both the $x$ and $y$ directions. Statistically-significant positive correlations ($p$-values labeled) between position and velocity for the blue points suggest expansion of the subcluster within the H\,{\sc ii} region.    
\label{fig:trifid_kin}}
\end{figure*}

When we examine the kinematics of Trifid Main, some substructure can be seen 
(Figure~\ref{fig:trifid_kin}). Cluster analysis in 4-dimensional $(x,y,v_x,v_y)$ space using Gaussian Mixture Models indicates two subclusters, with $\Delta\mathrm{BIC} = 42$ strongly supporting the 2-component model over a 1-component model.\footnote{This cluster solution is slightly different from that of \citet{Kuhn2014}. The explanation for this is that the present sample only includes optically visible sources with good Gaia astrometry, while \citet{Kuhn2014} included many more stars, but only identified clusters in 2 spatial dimensions. The densest portion of the Trifid cluster is near the center of the optical nebula near the location of HD 164492, but due to the pattern of extinction, our sample of Gaia sources is skewed toward the east.}
One of these subclusters corresponds to the densest part of Trifid Main and lies largely within the optically bright nebula (outline shown in pink). The other group includes stars around the northern periphery. 

The dense group within the optical nebula exhibits signs of expansion, although this effect is less apparent when both groups are lumped together. For the dense group (blue points in Figure~\ref{fig:trifid_kin}), Kendall's $\tau$ test -- a robust non-parametric test for correlation -- indicates weak positive correlation between position and velocity in the $x$ (east-west) direction ($p = 0.016$) and strong positive correlation between position and velocity in the $y$ (north-south) direction ($p = 3\times10^{-6}$).

To characterize this expansion, we use weighted ordinary least squares linear regression, with weights proportional to $1/\mathrm{error}^2$, and find mean velocity gradients of $0.4\pm0.2$~km~s$^{-1}$~pc$^{-1}$ for $v_x$ vs.\ $x$ and $1.3\pm0.2$~km~s$^{-1}$~pc$^{-1}$ for $v_y$ vs.\ $y$. However, these gradients only explain small fractions of the variance in velocity distributions, with coefficient of determination $R^2 = 0.02$ for the fit to $v_x$ vs.\ $x$ and $R^2 = 0.2$ for $v_y$ vs.\ $y$. Thus, we conclude that there is evidence for expansion, but most of the velocity dispersion is due to random motions. These velocity gradients are similar in magnitude to those observed in other star-forming regions like NGC~6530 (0.6~km~s$^{-1}$~pc$^{-1}$), Cep~B (1~km~s$^{-1}$~pc$^{-1}$), NGC~2244 (0.6~km~s$^{-1}$~pc$^{-1}$), IC~5070 (0.5~km~s$^{-1}$~pc$^{-1}$), and many others \citep[][]{Kuhn2019,Kuhn2020_nap,Wright2019}.

\subsection{Origin of the Trifid North Stars}

Trifid North stars are drifting northward and westward relative to the main cluster with mean velocities of $\langle v_x\rangle=0.8\pm0.3$~km~s$^{-1}$ and $\langle v_y\rangle =0.8\pm0.1$~km~s$^{-1}$. In projection, Trifid North is centered $\sim$7.2~pc north of the main cluster, so, if the Trifid North stars originated in the main cluster, the timescale for them to drift to their present location with their present velocities would be $\tau \approx 6$--14~Myr. Furthermore, the velocity dispersion for the stars in Trifid North $\sigma_\mathrm{1D} = 0.6$~km~s$^{-1}$ is even smaller than the group's mean velocity, implying that the Trifid North stars could not have individually drifted from the center of the main cluster to their present location in any less time. These timescales are considerably longer than the $\lesssim$1.5~Myr estimates for the ages of the young stars in Trifid \citep{Rho2008,Kalari2021} or the $\sim$1~Myr timescale for the hypothesized cloud-cloud collision \citep{Torii2011,Torii2017}. Thus, we conclude that the young stars in Trifid North formed separately from Trifid Main.

Given that the tangential velocities of the stars in Trifid North only differ from those of  Trifid Main by $<$1~km~s$^{-1}$, it seems likely that their radial velocities would also be similar. Unlike the line of sight to the Trifid nebula, where cloud components at both 2 and 8~km~s$^{-1}$ are found, the only low-velocity cloud detectable in the Trifid North field is one with emission peaked at $v_\mathrm{lsr}\sim 0$~km~s$^{-1}$, which appears to be a northern extension of the 2~km~s$^{-1}$ cloud (Figure~\ref{fig:sedigism}). 

\section{Discussion}\label{sec:discussion}

Our study has demonstrated that pre-main-sequence stars in both Trifid North and Trifid Main have statistically identical parallaxes and similar proper-motion distributions, indicating that they are both part of the same star-forming complex. Nevertheless, the kinematics of stars in Trifid North indicate that its stars could not have been born from the same part of the cloud as Trifid Main. Together, these regions span a total of $\sim$12~pc (in projection) and include areas of the star-forming complex with differing star and cloud properties. 

In Trifid Main there is a compact OB-dominated ``cluster'' that is kinematically expanding. This cluster is superimposed on an H\,{\sc ii} region and a molecular cloud ($>$10$^3$~$M_\odot$; Appendix~\ref{sec:cloud}) that appears to form an expanding shell. In contrast, Trifid North's members are spatially distributed and all low-mass; several massive stars projected within the same field are background nonmembers (Appendix~\ref{sec:massive_stars}). The stellar velocity dispersion in Trifid North is lower than in Trifid Main, and stars appear slightly older.  The cloud in the Trifid North region ($\sim$300~$M_\odot$; Appendix~\ref{sec:cloud}) is more diffuse and has a lower velocity dispersion, without the complex velocity structure seen in Trifid Main. Thus, the Trifid complex exhibits two different modes of star formation: one clustered with high-mass stars and the other distributed and composed of only low-mass stars. 

It is difficult to explain these results as star-formation triggered by a cloud-cloud collision. The alleged collision is between the 8 km~s$^{-1}$ cloud and the 2~km~s$^{-1}$ cloud in Trifid Main. However, the birth of stars in Trifid North, away from the site of the cloud-cloud collision, implies that a collision did not initiate star formation in the complex. An additional argument in favor of the cloud-cloud collision scenario was made by \citet{Kalari2021}, who find that their sample of H$\alpha$ emission and infrared-excess stars are concentrated around the edges of the H\,{\sc ii} region with a cavity in the center. They interpret this spatial distribution as tracing a ring of clouds compressed in the aftermath of a collision. However, this cavity was not seen in the X-ray selected probable cluster members, whose number density peaks near the center of the H\,{\sc ii} region \citep{Broos2013,Kuhn2015}. Thus, the deficit in H$\alpha$ emission and infrared-excess stars is likely the result of the difficulty of detecting such stars in regions with high optical and infrared nebulosity, like the center of Trifid Main \citep{2015ApJ...811...10R}.

The kinematics of the Trifid Main clouds can be alternatively explained by the expansion of the H\,{\sc ii} region, without requiring a cloud-cloud collision. Expanding H\,{\sc ii} regions have been observed in many star-forming regions powered by radiation pressure, thermal expansion, and/or stellar winds \citep{Tielens2010,FEEDBACK2020}. In Trifid Main, the diameter of the optical emission nebula is $\sim$3~pc in projection, so if this region corresponds to an H\,{\sc ii} bubble that began expanding $\sim$0.6~Myr ago \citep[the approximate ages of the O7 star HD~164492A~from][]{Petit2019}, then it would imply an average expansion velocity of $\sim$5~km~s$^{-1}$ for the bubble. This is comparable to the line-of-sight velocity difference between the approaching and receding cloud components. 

The situation in Trifid Main may be analogous to the expansion of Orion's Veil shell around the O7 star $\theta^1$~Ori~C. In Orion, the mass of the expanding shell has been calculated to be $1.5\times10^3$~$M_\odot$ (similar to the Trifid Main cloud), and it is expanding at $\sim$13~km~s$^{-1}$ \citep{Pabst2020}. This result from Orion suggests that the same expanding bubble mechanism would also be capable of producing the velocity differences observed in the Trifid Main cloud. 

The expansion of the cloud in Trifid Main appears to be reflected in the expansion of the group of stars projected within the boundary of the H\,{\sc ii} region. The expansion of a star cluster can be induced by changes to the gravitational potential as the expanding H\,{\sc ii} bubble drives gas out from within the cluster \citep{Tutukov1978,Hills1980,Adams2000,2019MNRAS.488.3406Z}. Thus, it appears plausible that that the observed expansion of both the cloud and cluster are causally connected. Furthermore, several protostars are embedded in the filaments making up the expanding shell around the Trifid Nebula \citep{2006ApJ...643..965R}, so if this shell is expanding, it is likely that these stars will be born with an outward velocity. 

In dramatic contrast to Trifid Main, in Trifid North both the cloud and stars have lower velocity distributions. The molecular cloud associated with Trifid North has only one velocity peak, and its range of velocities is much lower (Appendix~\ref{sec:cloud}). This is reflected in a stellar velocity dispersion that is much lower than in Trifid Main and much lower than in most other massive star-forming regions \citep[cf.][]{Kuhn2019}. Similarities between stellar and cloud kinematics, as seen here, could arise if stars inherit the kinematics of the gas from which they formed. 

Our results suggest that star-formation in the Trifid complex started as low-density ``distributed'' star formation, forming Trifid North, before advancing to ``clustered'' star formation in Trifid Main. 
This sequence matches theoretical models in which star formation commences with a low rate, but accelerates as material accretes along filaments from larger spatial scales into dense hubs where clustered star-formation takes place, leading to an increasing star-formation rate \citep{2014prpl.conf..291L,2018MNRAS.479.3254V,2019MNRAS.490.3061V}. The Trifid North region is drifting slowly away from the Trifid Main region. This is similar to results from other star-forming complexes where different subgroups of stars are observed to be dispersing rather than coalescing \citep{Kuhn2019,Kuhn2020_nap,2019A&A...626A..17C,2021ApJ...912..162P}. 

\section{Conclusions}\label{sec:conclusion}

The molecular cloud associated with the Trifid Nebula exhibits a complex velocity structure, with interacting components at 2 and 8~km~s$^{-1}$, indicating expansion that may be the aftermath of a hypothesized cloud-cloud collision. However, to the north of the optical nebula, we find only one cloud component at $\sim$0~km~s$^{-1}$, indicating that this section of the cloud would not have been affected by such a collision. To better understand how star-formation varies across this system, we obtained an X-ray observation of Trifid North to identify pre-main-sequence stars and compare their properties to the stars in Trifid Main. This complements a previous X-ray observation of Trifid Main. 

Our Chandra/ACIS-I X-ray observation of the Trifid North field covers a $\sim$34~pc$^2$ region 0.3$^\circ$ to north of Trifid Main. Out of 148 X-ray sources, 51 can be considered possible members, of which 13 sources with high-precision Gaia astrometry can be considered high-confidence members of the Trifid complex based on parallaxes and proper motions that are nearly identical to known members of the system ($d = 1180\pm25$~pc).  Their positions on the Gaia color-magnitude diagram are consistent with being pre-main-sequence stars, and infrared color-color diagrams suggest that these stars have extinctions ranging up to $A_V\sim10$~mag and that few have near-infrared excess. The positions of the Trifid North stars on the optical color-magnitude diagram suggest ages $\geq$3~Myr, which is older than the $\sim$1~Myr age estimate commonly assumed for Trifid Main \citep{Rho2008,Kalari2021}. Comparison between the X-ray luminosity functions for the stars in Trifid North and Trifid Main indicates that both Chandra X-ray observations have similar sensitivity limits in terms of X-ray luminosity. Based on this finding, we use scaling arguments to roughly estimate that the pre-main-sequence population in Trifid North is $\sim$9\% as rich as the population in Trifid Main. We have also provided a revised list of cluster members for Trifid Main using astrometry from Gaia EDR3, and we have recalculated distance-corrected X-ray properties for these stars. 

These results confirm that there is a pre-main-sequence stellar population spanning both the Trifid Main and Trifid North regions. The velocity dispersion for stars in Trifid North is only $\sim$0.6$\pm$0.2~km~s$^{-1}$ and their mean relative velocity along the direction of separation from Trifid Main is only $\sim$0.8$\pm$0.3~km~s$^{-1}$. Both of these imply that the stars could not have traveled to Trifid North from Trifid Main within the $\sim$1~Myr timescale for the formation of the main cluster. This implies that the stars in Trifid North likely formed from portions of the cloud outside the region with (potentially) interacting clouds, so such a cloud-cloud collision cannot be responsible for initiating star formation in this molecular cloud complex.

\begin{figure*}[t]
\centering
\includegraphics[angle=0.,width=0.95\textwidth]{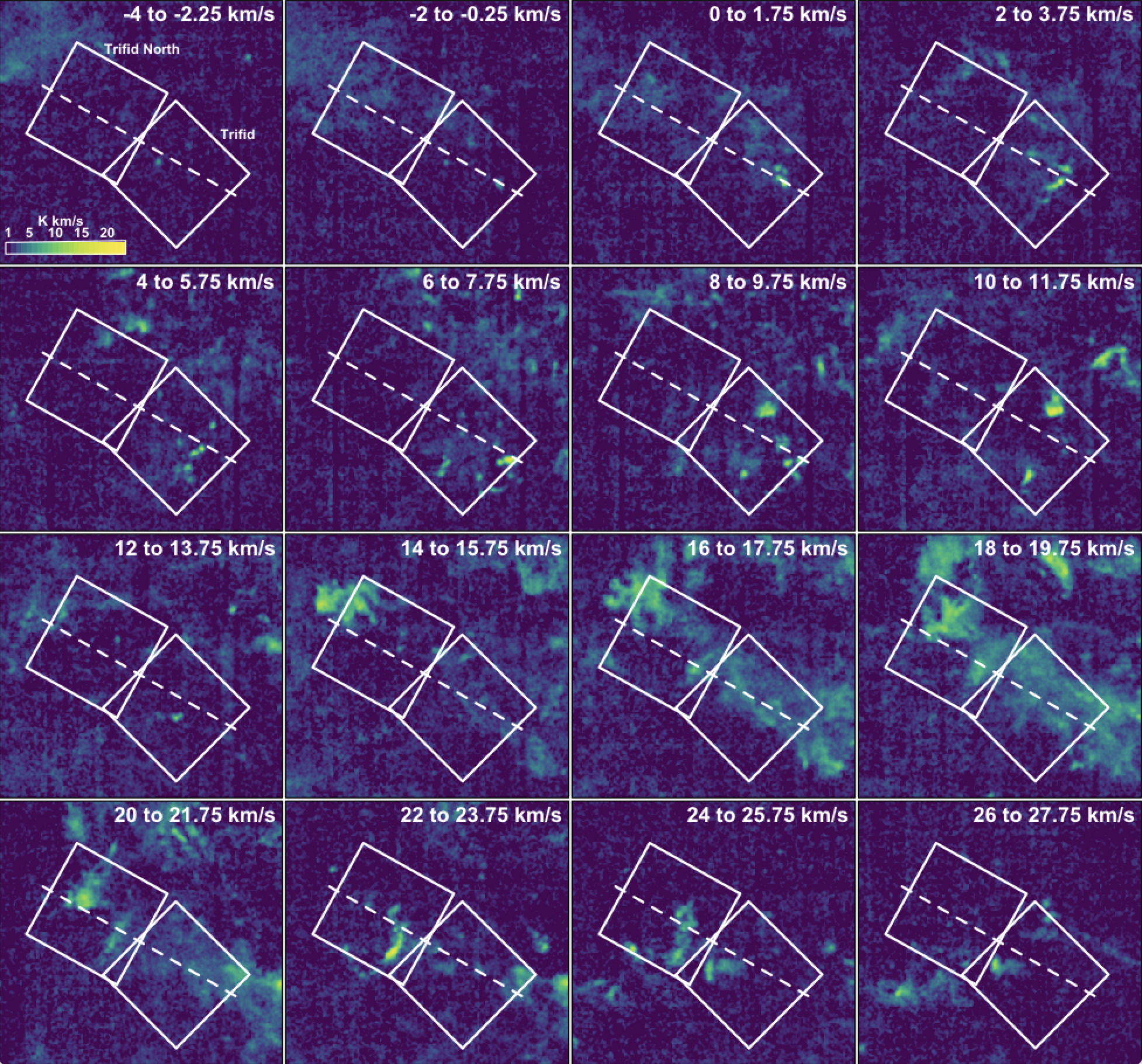}
\caption{Velocity slices of the $^{13}$CO $J=2$--1 SEDIGISM data cube. Each slice depicts emission integrated over 2~km~s$^{-1}$. The plots are in Galactic coordinates, with Galactic longitude $\ell$ increasing to the left and Galactic latitude $b$ increasing toward the top. The white squares are the Chandra fields of view, and the dashed line is the center of the slice in constant declination used in Figure~\ref{fig:sedigism}. The images are smoothed with a 9.5$^{\prime\prime}$ kernel to reduce noise. 
\label{fig:channel_map}}
\end{figure*}

An alternative mechanism to produce the kinematics of the molecular cloud is via the expansion of the H\,{\sc ii} region. The velocities required to create a bubble the size of the observed emission nebula are of similar scale to the velocities observed in the expansion of the molecular cloud around the main cluster. We find that among the stars making up Trifid Main, the group that lie within the H\,{\sc ii} bubble exhibits statistically significant signs of expansion, which could be a dynamical result of gas expulsion. Nevertheless, in either scenario, the cloud dynamics appear to have had an influence on the kinematics of the stars, creating a larger (expanding) velocity dispersion in the main cluster than in Trifid North.

The membership of young stars in the Trifid region is confused by the presence of two other stellar associations located behind Trifid. A loose grouping of stars (G7.4-0.2), identifiable due to their clustering in proper-motion space, spans the eastern edge of our Chandra Trifid North field, and a few of these stars are also detected in the X-ray. However, the estimated distance of 1310~pc places G7.4-0.2 behind Trifid. The second group is made up of the infrared-excess SPICY stars that are projected in the Trifid North field. These stars have parallaxes indicating that they are far behind the Trifid Nebula ($\gtrsim$2.7~pc). Furthermore, several newly confirmed massive stars projected in the Trifid North field also appear likely to be in the background (Appendix~\ref{sec:massive_stars}). It is possible that these YSOs and massive stars could be associated with the 18~km~s$^{-1}$ cloud (Appendix~\ref{sec:cloud}), whose kinematic distance would place it at the distance of the Scutum-Centaurus arm, which is consistent with the parallaxes of the stars in this group.

\appendix

\begin{figure*}[t]
\centering
\includegraphics[angle=0.,width=0.95\textwidth]{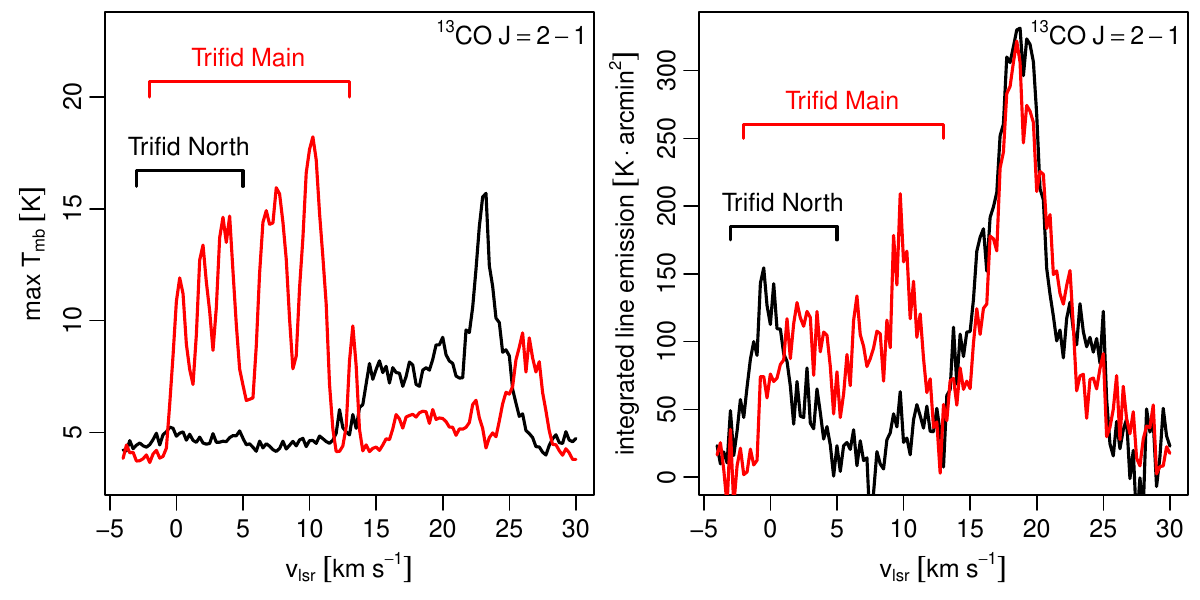}
\caption{$T_\mathrm{mb}$ versus $v_\mathrm{lsr}$ for $^{13}$CO $J=2$--1 in the Trifid North (black) and Trifid Main (red) fields. Left: The maximum $T_\mathrm{mb}$ at each velocity. Right: $T_\mathrm{mb}$ integrated over each field. 
\label{fig:co_velocity}}
\end{figure*}

\begin{figure*}[t]
\centering
\includegraphics[angle=0.,width=0.98\textwidth]{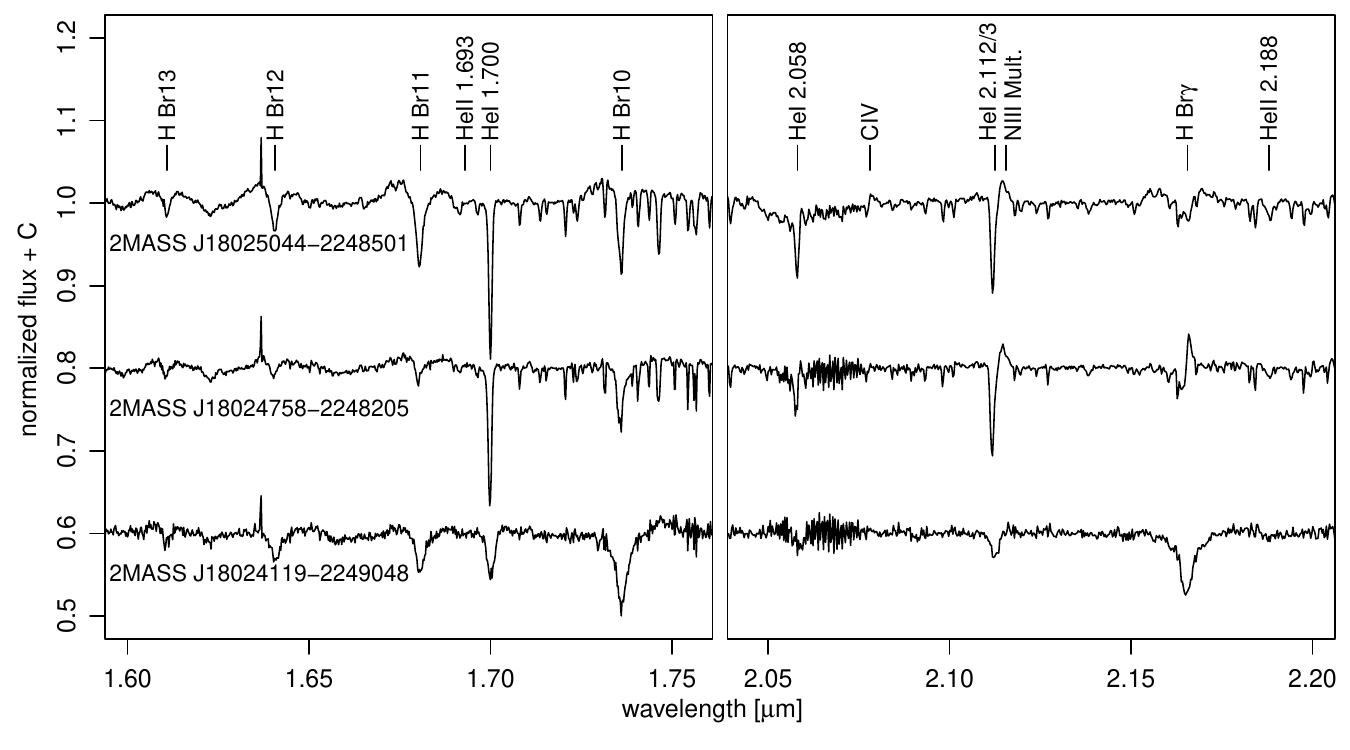}
\caption{Normalized ARCoIRIS spectra of the Trifid OB star candidates. Prominent OB star features used in the \citet{Hanson2005} classification scheme are marked. Several narrow lines, particularly around the Br10 line, can be attributed to imperfect atmospheric subtraction. 
\\
(The full spectra with wavelength ranges from $\sim$0.9--2.5~$\mu$m are included as ``data behind the figure.'') 
\label{fig:ob_arcoiris}}
\end{figure*}

\begin{figure*}[t]
\centering
\includegraphics[angle=0.,width=1\textwidth]{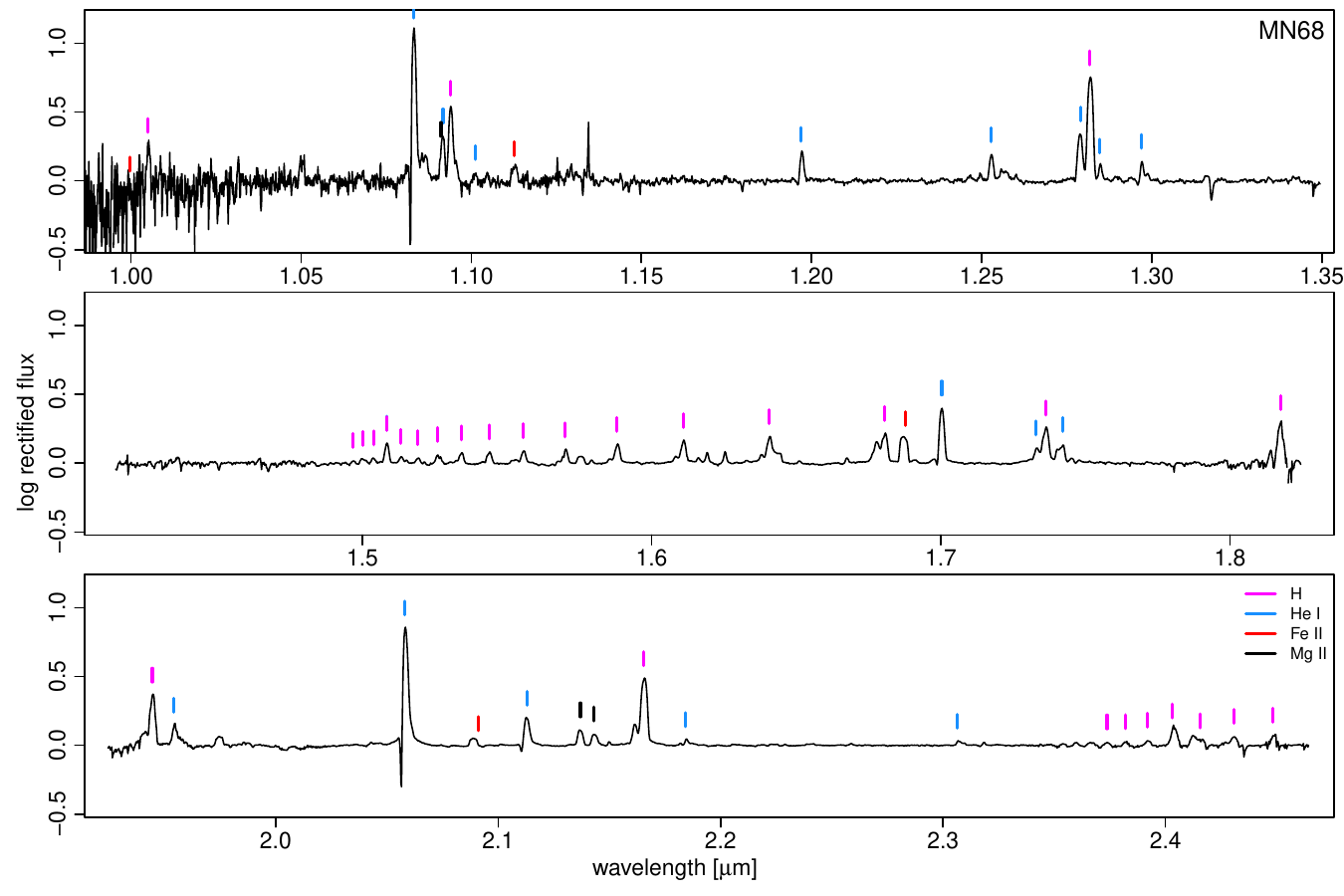}
\caption{Normalized near-infrared spectrum of MN68. Strong emission lines are detected from H (magenta tics) and He\,{\sc i} (light blue tics), with He\,{\sc i} exhibiting P~Cygni profiles. Emission lines from Fe\,{\sc ii} (red tics) and Mg\,{\sc ii} (black tics) are also marked.\\
(The reduced spectrum is provided in ``data behind the figure.'')  
\label{fig:mn68_spec}}
\end{figure*}

\section{Trifid Clouds in SEDIGISM}\label{sec:cloud}

Molecular line maps are available for the entire Trifid region in the $^{13}$CO $J=2$--1 transition (220.398684~GHz) from the Structure, Excitation, and Dynamics of the Inner Galactic Interstellar Medium \citep[SEDIGISM;][]{Schuller2021} survey using the Atacama Pathfinder Experiment (APEX) telescope. These data provide more complete coverage with higher velocity angular resolution (0.25~km~s$^{-1}$; $30^{\prime\prime}$) that complement the NANTEN2 $^{12}$CO $J=1$--0, $^{13}$CO $J=1$--0, $^{12}$CO $J=2$--1, and $^{13}$CO $J=2$--1 observations by \citet{Torii2011}.

Figure~\ref{fig:channel_map} shows emission from the $^{13}$CO $J=2$--1 transition integrated over sequential velocity ranges ($\Delta v= 2$~km~s$^{-1}$), between $v_\mathrm{lsr}=-4$~km~s$^{-1}$ and 28~km~s$^{-1}$. The Trifid North and Trifid Main fields used in our study are marked, as is the center of the slice in constant declination (dashed line) used in Figure~\ref{fig:sedigism} -- this slice is wide enough to cover nearly the entirety of the Chandra fields. 

Some $^{13}$CO emission is seen at almost any velocity in this range, but the emission does not necessarily originate from gas at the same distance. In the Galactic spiral arm model from \citet{Reid2016}, gas in the near Sagittarius Arm would have $v_\mathrm{lsr}\approx8$~km~s$^{-1}$ and gas in the near Scutum Arm would have $v_\mathrm{lsr}\approx18$~km~s$^{-1}$. \citet{Torii2017} have convincingly argued that gas between $\sim$0--13~km~s$^{-1}$ is associated with the Trifid Nebula based on its spatial distribution relative to the optical dust absorption features. They also suggest that the 18~km~s$^{-1}$ cloud is behind Trifid. Given that 18~km~s$^{-1}$ matches the expected velocity of the near Scutum Arm and some of the infrared-excess selected YSOs in the Trifid North field appear to be at $\sim$2.7~kpc (the approximate distance of the Scutum Arm), it seems plausible that CO emission in the velocity range $v_\mathrm{lsr}\approx13$--28~km~s$^{-1}$ comes from a molecular cloud far behind Trifid. This would leave the CO emission that peaks at $\sim$0~km~s$^{-1}$ (labeled on Figure~\ref{fig:sedigism}) as the best candidate for a cloud associated with the X-ray selected young stars in Trifid North. 

Figure~\ref{fig:co_velocity} shows properties of the emission as a function of velocity in both the Trifid North and Trifid Main fields of view. The maximum main-beam temperature, $T_\mathrm{mb}$, is different for the two regions (Figure~\ref{fig:co_velocity}, left panel). In Trifid Main, there is a high $\max T_\mathrm{mb}$ between 0~km~s$^{-1}$ and 13~km~s$^{-1}$ and a lower temperature beyond 13~km~s$^{-1}$. However this $\max T_\mathrm{mb}$ trend is reversed for Trifid North. The emission integrated over the fields of view (Figure~\ref{fig:co_velocity}, right panel) exhibits several peaks between 0~km~s$^{-1}$ and 13~km~s$^{-1}$ in Trifid Main, then one large peak at $\sim$18~km~s$^{-1}$. For Trifid North, we see only peaks at $\sim$0~km~s$^{-1}$ and at $\sim$18~km~s$^{-1}$. These plots suggest discontinuities at $\sim$13~km~s$^{-1}$ in both regions, reinforcing our hypothesis that the lower velocity clouds (0--13~km~s$^{-1}$) may be unrelated to the higher velocity clouds at $\sim$18~km~s$^{-1}$.

When estimating $^{13}$CO column densities, assumptions about excitation temperature $T_\mathrm{ex}$ can be a source of uncertainty. \citet{Torii2011} infer a temperature of $T_\mathrm{ex}=30$--50~K from ``large velocity gradient'' analysis based on the widths of multiple non-optically thick lines. A temperature of 30--50~K would be unusually high for a molecular cloud, and this is cited by \citet{Torii2011} as evidence for heating. The SEDIGISM $\max T_\mathrm{mb}$ values also indicate higher temperatures in at least some parts of Trifid Main. In the Trifid Main region, the peak temperature for the $^{13}$CO $J=2$--1 line is $T_\mathrm{mb}\approx18$~K (Figure~\ref{fig:co_velocity}, left), which implies a lower limit on excitation temperature of $T_\mathrm{ex}\gtrsim23$~K using Equation~88 from \citet{2015PASP..127..266M}.  

In Trifid Main, the multiple peaks between 0 and 13~km~s$^{-1}$ in both the $\max T_\mathrm{mb}$ and the integrated  $T_\mathrm{mb}$ plots (Figure~\ref{fig:co_velocity}) are compatible with the acceleration of dense cloudlets during the shredding of a molecular cloud by an expanding H\,{\sc ii} region. The observed velocities and structure are comparable to those obtained from numerical simulations of cloud disruption by radiative feedback \citep{2018MNRAS.477.5422A,2020MNRAS.497.3830F}. 

We estimate $^{13}$CO column densities from the SEDIGISM $^{13}$CO $J=2$--1 emission under the assumption of LTE using Equation~84 from \citet{2015PASP..127..266M}, then convert these to cloud masses assuming a $H_2$ to $^{13}$CO ratio of $3.8\times10^5$ \citep{2013ARA&A..51..207B}. 
Column densities are integrated within the Trifid North and Trifid Main fields of view, using a velocity range $-3$--5~km~s$^{-1}$ for Trifid North and $-2$--13~km~s$^{-1}$ for Trifid Main (Figure~\ref{fig:co_velocity}, right). Given the uncertainty in excitation temperature, we calculate cloud masses for several possible values, including $T_\mathrm{ex} = 12$~K (typical for cold molecular clouds), 23~K (obtained as a lower limit from the SEDIGISM data), and 40~K \citep[from][]{Torii2011}. For the Trifid North cloud, these give 400~$M_\odot$, 300~$M_\odot$, and 350~$M_\odot$, respectively. For Trifid Main, these give $>$830~$M_\odot$, 730~$M_\odot$, and 850~$M_\odot$, respectively. 
Without accurate estimates of $T_\mathrm{ex}$ it is difficult to determine whether parts of the Trifid Main cloud may be optically thick in the $^{13}$CO $J=2$--1 transition, so our mass estimates may be lower limits. The mass of the Trifid Main clouds derived here is lower than the total mass of $\sim$1.3--$3.2\times10^3$~$M_\odot$ calculated by \citet{Torii2011} from the NANTEN2 data. However, given the uncertainties, our values may not be inconsistent with theirs. 

\section{Newly Confirmed Massive Stars}\label{sec:massive_stars}

Several candidate massive stars are projected in the vicinity of Trifid North (Section~\ref{sec:introduction}). We have spectroscopically confirmed the massive nature of several of these objects, but, we argue, they are most likely located are located in the background of the Trifid Complex.  

Spectroscopic observations were made using the ARCoIRIS spectrograph on the Blanco Telescope at CTIO on March 21--22, 2016, which provided spectral coverage from 9,500--24,700~\AA\ at $R\approx 3500$ resolution. Observations were made with a A-B-B-A nodding pattern, with exposure times for individual frames ranging from 10~s to 30~s. Data were reduced using a modified version of Spextool v4.1 \citep{spextool}, using the A0 star HD~159415 for telluric correction and flux calibration. 

\subsection{OB Stars}\label{sec:ob}

\citet{2017ApJ...838...61P} identified four candidate OB in the Trifid region (Figure~\ref{fig:4panel}) based on X-ray and mid-infrared photometry. Three of these are clustered in a small group to the northeast of the Trifid Nebula, within the region of overlap between the Trifid Main and Trifid North Chandra fields. These three stars, 2MASS J18025044-2248501 ($H = 8.0$~mag), 2MASS J18024758-2248205 ($H=8.3$~mag), 2MASS J18024119-2249048 ($K_s = 9.7$~mag), were observed for 3.7~min, 4~min, and 6~min with air masses of 1.2, 1.13, and 1.06, respectively. The spectral region used for classification is shown in Figure~\ref{fig:ob_arcoiris}.

Both 2MASS J18025044-2248501 and 2MASS J18024758-2248205 have detectable absorption in the He\,{\sc ii} 1.693~$\mu$m and 2.188~$\mu$m lines and emission in a feature at 2.1155~$\mu$m identified by \citet{Hanson2005} as a N\,{\sc iii} multiplet. These features are not present in stars from the \citet{Hanson2005} library with spectral types O9.5 or later, but the relative weakness of these features compared to nearby He\,{\sc i} lines suggests a spectral type of O8 or later. Thus, we classify 2MASS J18025044-2248501 and 2MASS J18024758-2248205 as O8-9. 

2MASS J18024119-2249048 shows a possible hint of He\,{\sc ii} emission at 1.693~$\mu$m, but lacks  N\,{\sc iii} emission at 2.1155~$\mu$m. The He\,{\sc I} 1.700~$\mu$m absorption line is slightly stronger than the H Br11 line and the He\,{\sc ii} 2.112/3~$\mu$m feature is about one third the strength of the H Br$\gamma$ absorption line, both of which best match the example O9.5 spectrum from \citet{Hanson2005}. The strength of the H\,{\sc i} lines rules out spectral types of B2 or later. Thus, we classify  J18024119-2249048 as O9.5-B1.

Of these stars, only 2MASS J18024758-2248205 has a good RUWE, indicating that its parallax can be considered reliable. The Gaia EDR3 parallax for this star is $0.2269\pm0.0245$, and the parallax zero-point correction is $-0.0169$~mas, yielding a corrected parallax of $0.2437$~mas. This parallax places this O star well behind the Trifid Nebula and is inconsistent with the parallax of either Trifid North or Trifid Main at $>$5$\sigma$. Given that the other 2MASS J18025044-2248501 and 2MASS J18024119-2249048 are part of the same subcluster in angular position, it seems likely that these stars are in the background too.

\subsection{Evolved Massive Star}\label{sec:wr}

We obtained an ARCoIRIS spectrum of [GKF2010] MN68 (18:02:22.340 $-$22:38:00.24; $J=13.6$~mag, $H=11.1$~mag, $K_s = 9.6$~mag) with a total integration time of 22~min at an airmass of 1.28. This source, which lies near the center of the Trifid North field, was identified as a candidate evolved massive star on the basis of a circumstellar nebula seen in Spitzer/IRAC images \citep{Gvaramadze2010,Wachter2010}. The star is not optically visible, so it has no Gaia astrometry. Furthermore, it is not detected by Chandra and has an upper limit of $F_X \leq 3\times10^{-15}$~erg~s$^{-1}$ in the 0.5--8.0~keV band. \citet{2015MNRAS.452.2858K} report a spectral classification of B[e]/LBV but do not show the spectrum. 

The star's spectrum (Figure~\ref{fig:mn68_spec}) is dominated by emission lines from hydrogen and neutral helium. The He\,{\sc i} lines have P Cygni profiles, with full widths at half maximum (FWHM) of $\sim$300~km~s$^{-1}$. Emission lines from Mg\,{\sc ii} and Fe\,{\sc ii} are also detected. However, no He\,{\sc ii} lines are clearly observed. 

The spectrum closely resembles the near-infrared spectra of B-type supergiants like HDE 316285 \citep{Hillier1998} or IGR J16318–4848 \citep{Fortin2020}. In particular, the H, He\,{\sc i}, and Mg\,{\sc ii} line shapes in the wavelength ranges from 1.08--1.10~$\mu$m \citep[compare to][their Figure~3d]{Hillier1998} and between 2.04--2.18~$\mu$m \citep[compare to][their Figure~2]{Hillier1998} resemble HDE 316285. Furthermore, the Fe\,{\sc ii} lines in the MN68 spectrum tend to have flat tops, similar to the the Fe\,{\sc ii} lines in both HDE 316285 and IGR J16318–4848, which \citet{Fortin2020} attribute to a disk wind for IGR J16318–4848. 

The red $J-H$ color of MN68 suggests an extinction of $A_J \approx 6.6$~mag, yielding a de-reddened $J$-band magnitude of $\sim$7~mag. Given the high luminosities of B-type supergiants, if MN68 were a supergiant, it would imply a distance modulus that places the star far behind the Trifid Nebula. 

\section{Corrections to X-ray Median Energy}\label{sec:obf}

To estimate the shift in median energy between early in the Chandra mission and 2019 due to accumulation of contaminants on the ACIS detector filter, we use simulated X-ray spectra generated by the XSPEC {\it fakeit} tool. Changes in quantum efficiency are encoded in the response matrix files (RMFs) and ancillary response files (ARFs), which we obtained from both a 2003 observation of NGC~2362 (ObsID~4469) and from the current observation. We simulated absorbed thermal plasma models \citep{1983ApJ...270..119M,2001ApJ...556L..91S}, with temperature components taken from fits to a sample of real pre-main-sequence stars \citep{Getman2005}, and a grid of absorbing column densities from $N_\mathrm{H}=0$~cm$^{-2}$ to $2\times10^{23}$~cm$^{-2}$. For the full band, the mean shift ranges from 0.05~keV to 0.3~keV ($\pm$0.05~keV uncertainty) as a function of observed median energy, and, in the hard band, the mean shift ranges from 0.0~keV to 0.06~keV ($\pm$0.01~keV uncertainty). 

\vspace{3.5mm}
\noindent This research was supported by Chandra grant GO9-20002X. This work is based on data from ESA’s Gaia mission \citep{2016arXiv160904172G}, processed by the Data Processing and Analysis Consortium, funded by national institutions, particularly those participating in the Gaia Multilateral Agreement. We thank David James and Sean Points for assistance with ARCoIRIS, Katelyn Allers for the modified Spextool software, Paul Crowther and Philip Massey for suggestions about spectral classification, and the anonymous referee for useful comments. I.E.F.\ was supported by Caltech’s Freshman Summer Research Institute (FSRI). M.G.\ is supported by the EU Horizon 2020 research and innovation programme under grant agreement No.\ 101004719.

\facility{2MASS, APEX, Blanco (ARCoIRIS), CXO (ACIS), Gaia, Herschel, IRSA, Spitzer (IRAC, MIPS), UKIRT} 

\software{
          ACIS Extract \& TARA \citep{Broos2010,2012ascl.soft03001B},
          AstroLib \citep{1993ASPC...52..246L},
          astropy \citep{cite-astropy13, cite-astropy18},
          CIAO \citep{2006SPIE.6270E..1VF},
          HEASOFT \citep{2014ascl.soft08004N},
          MARX \citep{2013ascl.soft02001W},
          mclust \citep{mclust5},
          numpy \citep{2011CSE....13b..22V},
          PIMMS \citep{1993Legac...3...21M},
          pvextractor \citep{RadioAstronomyToolsInPython},
          R \citep{RCoreTeam2018}, 
          SAOImage DS9 \citep{2003ASPC..295..489J},
          scipy \citep{2020NatMe..17..261V},
          Spextool \citep{spextool},
          TOPCAT \& STILTS \citep{2005ASPC..347...29T},
          wavdetect \citep{2002ApJS..138..185F},
          weights \citep{weights2016},
          XPHOT \citep{Getman2010},
          XSPEC \citep{1996ASPC..101...17A}
          }

\bibliography{ms.bbl}

\begin{deluxetable*}{lrrrrrrrrrrrrrr}
\tablecaption{X-ray source properties (Trifid North)\label{tab:xray}}
\tabletypesize{\tiny}\tablewidth{0pt}\rotate
\tablehead{
\colhead{CXOU} & 
\colhead{R.A.} &
\colhead{Decl.} &
\colhead{Error} &
\colhead{$SC_t$} &
\colhead{$SC_h$} &
\colhead{$NC_t$} &
\colhead{$NC_h$} &
\colhead{$\theta$} &
\colhead{$P_B$} &
\colhead{$P_\mathrm{KS}$} &
\colhead{$ME_\mathrm{2019}$} &
\colhead{$ME_\mathrm{corr}$} &
\colhead{$\log F_t$} &
\colhead{$\log F_h$}\\
\colhead{} & 
\colhead{(ICRS)} &
\colhead{(ICRS)} &
\colhead{(arcsec)} &
\colhead{(counts)} &
\colhead{(counts)} &
\colhead{(counts)} &
\colhead{(counts)} &
\colhead{(arcmin)} &
\colhead{} &
\colhead{} &
\colhead{(keV)} &
\colhead{(keV)} &
\colhead{(erg s cm$^{-1}$)} &
\colhead{(erg s cm$^{-1}$)}\\
\colhead{(1)} & 
\colhead{(2)} &
\colhead{(3)} &
\colhead{(4)} &
\colhead{(5)} &
\colhead{(6)} &
\colhead{(7)} &
\colhead{(8)} &
\colhead{(9)} &
\colhead{(10)} &
\colhead{(11)} &
\colhead{(12)} &
\colhead{(13)} &
\colhead{(14)} &
\colhead{(15)}
  }
\startdata
  J180235.9-224315 & 270.6496502 & -22.72084404 & 0.313 & 13 & 8 & 11.9 & 7.1 & 4.4 & $3\times10^{-10}$ & $6\times10^{-1}$ & 2.07 & 1.78 & -14.41$\pm$0.18 & -14.49$\pm$0.22\\
  J180236.4-223935 & 270.6518444 & -22.65993954 & 0.153 & 24 & 11 & 23.5 & 10.6 & 3.6 & $4\times10^{-32}$ & $2\times10^{-6}$ & 1.89 & 1.6 & -14.2$\pm$0.11 & -14.29$\pm$0.17\\
  J180236.5-224410 & 270.6520955 & -22.73630954 & 0.328 & 20 & 14 & 18.2 & 12.7 & 5.1 & $1\times10^{-14}$ & $5\times10^{-1}$ & 2.64 & 2.38 & -14.09$\pm$0.14 & -14.11$\pm$0.16\\
  J180236.5-224054 & 270.6523536 & -22.68174294 & 0.191 & 17 & 15 & 16.5 & 14.6 & 3.5 & $1\times10^{-20}$ & $1\times10^{-1}$ & 3.58 & 3.5 & -13.95$\pm$0.13 & -13.99$\pm$0.14\\
  J180237.1-224742 & 270.6548669 & -22.79508834 & 0.596 & 38 & 15 & 29.1 & 8.2 & 8.1 & $1\times10^{-11}$ & $1\times10^{-2}$ & 1.5 & 1.27 & -14.07$\pm$0.11 & -14.25$\pm$0.24\\
  J180237.5-223816 & 270.6563687 & -22.63783074 & 0.101 & 80 & 46 & 79.0 & 45.2 & 4.3 & 0 & 0 & 2.31 & 2.05 & -13.5$\pm$0.06 & -13.6$\pm$0.08\\
  J180238.0-223458 & 270.6586381 & -22.58291854 & 0.398 & 39 & 29 & 35.6 & 26.4 & 6.7 & $4\times10^{-26}$ & $6\times10^{-3}$ & 3.29 & 3.18 & -13.69$\pm$0.09 & -13.77$\pm$0.11\\
  J180238.3-223751 & 270.6599922 & -22.63096834 & 0.329 & 15 & 12 & 14.0 & 11.2 & 4.7 & $5\times10^{-13}$ & $1\times10^{-1}$ & 4.17 & 4.11 & -13.93$\pm$0.15 & -14.01$\pm$0.16\\
  J180238.8-223223 & 270.6618095 & -22.53979434 & 0.178 & 257 & 120 & 246.8 & 113.0 & 9.0 & 0.0 & 0.0 & 1.89 & 1.6 & -13.12$\pm$0.04 & -13.22$\pm$0.05\\
  J180238.8-223527 & 270.6619471 & -22.59102204 & 0.373 & 38 & 36 & 34.4 & 33.3 & 6.4 & $2\times10^{-25}$ & $2\times10^{-1}$ & 4.53 & 4.48 & -13.51$\pm$0.09 & -13.58$\pm$0.09
\enddata
\tablecomments{Properties of 143 X-ray sources in the Trifid North field. Column 1: Source name. Columns 2--3: Source coordinates. Column 4: Positional uncertainty. Columns 5--6: Number of counts in the extraction aperture for the total (0.5--8.0~keV) and hard (2--8~keV) bands. Columns 7--8: Background-subtracted net counts in the total and hard bands. Column 9: Off-axis angle. Column 10: Probability of producing the observed source counts as a fluctuation of the background level. Column 11: Probability of producing the observed Kolmogorov--Smirnov test statistic from a non-variable source (values close to zero indicate light-curve variability). Columns 12: Median energy (total band) as observed in 2019. Column 13: Median energy (total band) with the correction from Appendix~\ref{sec:obf}. Columns 14--15: X-ray flux in the total and hard bands calculated using Equations~8--9 in \citet{Broos2010}.\\ \\
(This table is available in its entirety in a machine-readable form in the online journal. A portion is shown here for guidance regarding its form and content.)
}
\end{deluxetable*}

\begin{deluxetable*}{rcl}
\tablecaption{X-ray Selected Trifid North Member Candidates\label{tab:mem}}
\tabletypesize{\tiny}\tablewidth{0pt}
\tablehead{
  \colhead{Column} &  \colhead{Column ID}   &  \colhead{Description} 
}
\startdata
1 & CXOU & X-ray Source Designation\\
2 & $\log N_\mathrm{H}$ & X-ray absorption as hydrogen column density\\
3 & $e\_\log N_\mathrm{H}$ & error on $N_H$\\
4 & $\log L_X$ & absorption-corrected luminosity in the 0.5--8.0~keV band \\
5 & $e\_\log L_X$ & error on $L_X$ \\
6 & f\_kin & Flag indicating kinematic membership$^\mathrm{a}$\\
\hline
\multicolumn{3}{c}{Gaia Columns}\\
7 & GaiaEDR3 & Gaia EDR3 source designation\\
8 & RA\_GaiaEDR3 & ICRS R.A.\ of the Gaia counterpart\\
9 & Dec\_GaiaEDR3 & ICRS decl.\ of the Gaia counterpart\\
10 & parallax\_GaiaEDR3 & Gaia EDR3 parallax\\
11 & e\_parallax\_GaiaEDR3 & error on Gaia EDR3 parallax\\
12 & zp & Gaia DR3 parallax zero point estimate\\
13 & parallax\_corr & Zero-point corrected parallax\\
14 & pmra\_GaiaEDR3 & Gaia EDR3 proper motion in R.A.\\
15 & e\_pmra\_GaiaEDR3 & error on Gaia EDR3 proper motion in R.A.\\
16 & pmdec\_GaiaEDR3 & Gaia EDR3 proper motion in decl.\\
17 & e\_pmdec\_GaiaEDR3 & error on Gaia EDR3 proper motion in  decl.\\
18 & G & Gaia EDR3 $G$-band magnitude.\\
19 & BP & Gaia EDR3 $G_\mathrm{BP}$-band magnitude\\
20 & RP & Gaia EDR3 $G_\mathrm{RP}$-band magnitude\\
21 & RUWE & Gaia EDR3 renormalized unit weight error\\
22 & aen & Gaia EDR3 astrometric excess noise\\
23 & sigma5d\_max & semimajor axis of the Gaia EDR3 astrometric error ellipse\\
\hline
\multicolumn{3}{c}{UKIDSS Columns$^\mathrm{b}$}\\
24 & RA\_UKIDSS & ICRS R.A.\ of the UKIDSS counterpart\\
25 & Dec\_UKIDSS & ICRS decl.\ of the UKIDSS counterpart\\
26 & $J$ & UKIDSS $J$-band magnitude\\
27 & e\_$J$ & Error on $J$-band magnitude\\
28 & f\_$J$ & UKIDSS $J$-band magnitude flag\\
29 & $H$ & UKIDSS $H$-band magnitude\\
30 & e\_$H$ & Error on $H$-band magnitude\\
31 & f\_$H$ & UKIDSS $H$-band magnitude flag\\
32 & $K$ & UKIDSS $K_s$-band magnitude\\
33 & e\_$K$ & Error on $K_s$-band magnitude\\
34 & f\_$K$ & UKIDSS $K_s$-band magnitude flag\\
\hline
\multicolumn{3}{c}{Other Cross-Matched Catalogs}\\
35 & 2MASS & 2MASS source designation\\
36 & Spitzer & Spitzer source designation\\
\enddata
\tablecomments{ 
Properties of 51 X-ray sources identified as members or candidate members of Trifid North. In addition to the quantities derived in this paper, for the convenience of the user this table also provides select columns from the Gaia EDR3 and UKIDSS catalogs. 
\\
(This table is available in its entirety in a machine-readable form in the online journal. The list of columns is shown here for guidance regarding its form and content.)
}
\vspace*{-0.05in}
\tablenotetext{$a$}{Kinematic member classes include: ``A'' -- source is a high-confidence member of the Trifid Nebula region; ``B'' -- source is a member of the group likely associated with G7.4-0.2; ``C'' -- source is likely to be an unrelated field star. Objects without flags lack Gaia astrometry of sufficient precision to determine kinematic group.}
\vspace*{-0.05in}
\tablenotetext{$b$}{UKIDSS sources in \citet{King2013} are not given designations, so coordinates are used to identify them instead. Only photometry with ``OO'' (good) or ``OV'' (variable) flags is included. }
\end{deluxetable*}

\begin{deluxetable*}{rcl}
\tablecaption{Revised members of Trifid Main\label{tab:trifid}}
\tabletypesize{\tiny}\tablewidth{0pt}
\tablehead{
  \colhead{Column} &  \colhead{Column ID}   &  \colhead{Description} 
}
\startdata
1 & MPCM & IAU Source Designation\\
2 & RA\_MPCM &  R.A.\ (J2000) from \citet{Broos2013} \\
3 & Dec\_MPCM & Dec.\ (J2000) from \citet{Broos2013}\\
4 & $\log N_\mathrm{H}$ & X-ray absorption as hydrogen column density\\
5 & $e\_\log N_\mathrm{H}$ & error on $\log N_H$\\
6 & $\log L_X$ & absorption-corrected luminosity in the 0.5--8.0~keV band \\
7 & $e\_\log L_X$ & error on $\log L_X$ \\
8 & f\_kin & Flag indicating high-confidence kinematic membership\\
\hline
\multicolumn{3}{c}{Gaia Columns}\\
9 & GaiaEDR3 & Gaia EDR3 source designation\\
10 & ra\_GaiaEDR3 & ICRS R.A.\ of the Gaia counterpart\\
11 & dec\_GaiaEDR3 & ICRS decl.\ of the Gaia counterpart\\
12 & parallax\_GaiaEDR3 & Gaia EDR3 parallax\\
13 & e\_parallax\_GaiaEDR3 & error on Gaia EDR3 parallax\\
14 & zp & Gaia DR3 parallax zero point estimate\\
15 & parallax\_corr & Zero-point corrected parallax\\
16 & pmra\_GaiaEDR3 & Gaia EDR3 proper motion in R.A.\\
17 & e\_pmra\_GaiaEDR3 & error on Gaia EDR3 proper motion in R.A.\\
18 & pmdec\_GaiaEDR3 & Gaia EDR3 proper motion in decl.\\
19 & e\_pmdec\_GaiaEDR3 & error on Gaia EDR3 proper motion in  decl.\\
20 & G & Gaia EDR3 $G$-band magnitude.\\
21 & BP & Gaia EDR3 $G_\mathrm{BP}$-band magnitude\\
22 & RP & Gaia EDR3 $G_\mathrm{RP}$-band magnitude\\
23 & ruwe & Gaia EDR3 renormalized unit weight error\\
24 & aen & Gaia EDR3 astrometric excess noise\\
25 & sigma5d\_max & semimajor axis of the Gaia EDR3 astrometric error ellipse\\
\hline
\multicolumn{3}{c}{UKIDSS Columns}\\
26 & ra\_UKIDSS & ICRS R.A.\ of the UKIDSS counterpart\\
27 & dec\_UKIDSS & ICRS decl.\ of the UKIDSS counterpart\\
28 & $J$ & UKIDSS $J$-band magnitude\\
29 & e\_$J$ & Error on $J$-band magnitude\\
30 & f\_$J$ & UKIDSS $J$-band magnitude flag\\
31 & $H$ & UKIDSS $H$-band magnitude\\
32 & e\_$H$ & Error on $H$-band magnitude\\
33 & f\_$H$ & UKIDSS $H$-band magnitude flag\\
34 & $K$ & UKIDSS $K_s$-band magnitude\\
35 & e\_$K$ & Error on $K_s$-band magnitude\\
36 & f\_$K$ & UKIDSS $K_s$-band magnitude flag\\
\hline
\multicolumn{3}{c}{Other Cross-Matched Catalogs}\\
37 & 2MASS & 2MASS source designation\\
38 & Spitzer & Spitzer source designation\\
\enddata
\tablecomments{ 
Candidate members of Trifid Main with re-evaluated membership classifications and revised X-ray properties using the updated distance estimate of 1180~pc. Out of the 532 candidate from \citet{Broos2013}, 17 are rejected as nonmembers (not included in this table) and 515 are retained as possible members (listed here), 149 of which are deemed high-confidence members (i.e., labeled ``A'' in Column~8). \\
(This table is available in its entirety in a machine-readable form in the online journal. The list of columns is shown here for guidance regarding its form and content.)
}
\end{deluxetable*}

\begin{deluxetable*}{lrrrrrrrr}
\tablecaption{Gaia members of G7.4-0.2\label{tab:groupB}}
\tabletypesize{\small}\tablewidth{0pt}
\tablehead{
  \colhead{Gaia EDR3} &
  \colhead{R.A.} &
  \colhead{Decl.} &
  \colhead{$\varpi$} &
  \colhead{$\mu_{\alpha^\star}$} &
  \colhead{$\mu_\delta$} &
  \colhead{G} &
  \colhead{BP} &
  \colhead{RP}\\
  \colhead{} &
  \colhead{(deg)} &
  \colhead{(deg)} &
  \colhead{(mas)} &
  \colhead{(mas yr$^{-1}$)} &
  \colhead{(mas yr$^{-1}$)} &
  \colhead{(mag)} &
  \colhead{(mag)} &
  \colhead{(mag)}\\
  \colhead{(1)} &
  \colhead{(2)} &
  \colhead{(3)} &
  \colhead{(4)} &
  \colhead{(5)} &
  \colhead{(6)} &
  \colhead{(7)} &
  \colhead{(8)} &
  \colhead{(9)}
  }
\startdata
4069248313428288128 & 270.34179 & $-$23.04160 & 0.720$\pm$0.280 & 0.840$\pm$0.280 & $-$2.763$\pm$0.216 & 18.8 & 20.0 & 17.8\\
4069468318822935936 & 270.35245 & $-$22.70212 & 0.716$\pm$0.181 & 0.886$\pm$0.191 & $-$2.685$\pm$0.134 & 18.0 & 19.4 & 16.8\\
4069468147029582720 & 270.37349 & $-$22.70705 & 0.746$\pm$0.177 & 0.772$\pm$0.165 & $-$2.844$\pm$0.119 & 17.8 & 19.4 & 16.6\\
4069248249012818304 & 270.37375 & $-$23.03059 & 0.755$\pm$0.030 & 0.420$\pm$0.033 & $-$2.360$\pm$0.023 & 14.9 & 15.8 & 14.0\\
4069467081872282752 & 270.37895 & $-$22.76373 & 0.819$\pm$0.138 & 0.789$\pm$0.084 & $-$2.482$\pm$0.063 & 16.8 & 18.0 & 15.7\\
4069270892081677184 & 270.39697 & $-$23.03388 & 0.652$\pm$0.145 & 0.616$\pm$0.179 & $-$2.474$\pm$0.129 & 17.9 & 19.3 & 16.8\\
4069865693489953792 & 270.47628 & $-$22.43080 & 0.786$\pm$0.227 & 0.873$\pm$0.289 & $-$2.646$\pm$0.206 & 18.8 & 20.3 & 17.6\\
4069242923253269120 & 270.48040 & $-$23.15074 & 0.735$\pm$0.141 & 0.774$\pm$0.171 & $-$2.389$\pm$0.123 & 17.9 & 18.9 & 16.7\\
4069473511453492352 & 270.48707 & $-$22.67485 & 0.849$\pm$0.095 & 0.552$\pm$0.077 & $-$2.715$\pm$0.053 & 16.8 & 17.8 & 15.8\\
4069470453441642496 & 270.49381 & $-$22.71172 & 0.746$\pm$0.020 & 0.503$\pm$0.019 & $-$2.436$\pm$0.013 & 10.8 & 11.3 & 10.1\\
\enddata
\tablecomments{Gaia sources selected as members of G7.4-0.2 (Group B). Column 1: Source designation. Columns 2--3: Coordinates in the ICRS system. Column 4: Zero-point corrected parallax. Columns 5--6: Proper motions. Columns 7--8: Gaia photometric magnitudes.
}
\end{deluxetable*}

\end{document}